\documentclass[twocolumn]{aastex63}

\usepackage{graphicx}
\usepackage{color, colortbl}
\usepackage{float}

\usepackage{lineno}
\shorttitle{XTE J2012+381}
\shortauthors{Draghis et al.}

\begin{document}

\title{An Extreme Black Hole in the Recurrent X-ray Transient XTE J2012+381}

\author[0000-0002-2218-2306]{Paul A. Draghis}
\email{pdraghis@umich.edu}
\affiliation{Department of Astronomy, University of Michigan, 1085 South University Avenue, Ann Arbor, MI 48109, USA}

\author[0000-0003-2869-7682]{Jon M. Miller}
\affiliation{Department of Astronomy, University of Michigan, 1085 South University Avenue, Ann Arbor, MI 48109, USA}

\author[0000-0002-4024-6967]{McKinley C. Brumback}
\affiliation{Department of Astronomy, University of Michigan, 1085 South University Avenue, Ann Arbor, MI 48109, USA}

\author[0000-0002-9378-4072]{Andrew C. Fabian}
\affiliation{Institute of Astronomy, University of Cambridge, Madingley Road, Cambridge CB3 OHA, UK}

\author[0000-0001-5506-9855]{John A. Tomsick}
\affiliation{Space Sciences Laboratory, 7 Gauss Way, University of California, Berkeley, CA, 94720-7450, USA}

\author[0000-0002-0572-9613]{Abderahmen Zoghbi}
\affiliation{Department of Astronomy, University of Maryland, College Park, MD, 20742, USA}
\affiliation{HEASARC, Code 6601, NASA/GSFC, Greenbelt, MD, 20771, USA}
\affiliation{CRESST II, NASA Goddard Space Flight Center, Greenbelt, MD, 20771, USA}

\begin{abstract}
The black hole candidate XTE J2012+381 underwent an outburst at the end of 2022. We analyzed 105 NICER observations and 2 NuSTAR observations of the source during the outburst. The NuSTAR observations of the $M \sim10M_\odot$ black hole indicate clear signs of relativistic disk reflection, which we modeled to measure a BH spin of $a=0.988^{+0.008}_{-0.030}$ and an inclination of $\theta=68^{+6}_{-11}$ degrees ($1\sigma$ statistical errors). In our analysis, we test an array of models and examine the effect of fitting NuSTAR spectra alone versus fitting simultaneously with NICER. We find that when the underlying continuum emission is properly accounted for, the reflected emission is similarly characterized by multiple models. We combined 52 NICER spectra to obtain a spectrum with an effective exposure of 190 ks in order to probe the presence of absorption lines that would be suggestive of disk winds, but the resulting features were not statistically significant. We discuss the implications of this measurement in relation to the overall BH spin distribution in X-ray binary systems.
\end{abstract}

%\keywords{accretion, accretion disks -- black hole physics -- individual (XTE J2012+381) -- X-rays: binaries}

\section{Introduction} \label{sec:intro}
The observed spin distribution of stellar mass black holes (BHs) in X-ray binary (XB) systems is in disagreement with the spin distribution of BHs in merging binary black hole (BBH) systems observed through gravitational waves (GW), with BHs in XB having preferentially high spins, whereas BHs in BBHs have preferentially low spins (\citealt{2022ApJ...929L..26F, 2023ApJ...946...19D}). It is important however to acknowledge that while the distribution of spins in BBH accounts for selection effects and observational biases, the distribution of spins of BHs in XB is built based only on the observed spins and there may be unknown selection effects. Furthermore, the measured spin values are reported with only statistical uncertainties, as the systematic uncertainties are not yet well understood. The most pragmatic approach to quantifying the systematic uncertainties of the spin measurements of BHs in XBs and to attempt to quantify the observational biases is to measure the BH spin in as many sources as possible. 

%paragraph about relativistic reflection and NuSTAR
For BHs in XBs, the preferred spin measurement techniques that use of X-ray spectroscopy are the ``continuum fitting" method (see, e.g., \citealt{2009ApJ...701.1076G, 2023MNRAS.520.5803F}) and the ``relativistic reflection" method (see, e.g., \citealt{2000PASP..112.1145F, 2006ApJ...652.1028B, 2007ARA&A..45..441M, 2010ApJ...724.1441M, 2020ApJ...900...78D}). Both methods come with a series of assumptions and simplifications. For a review of the efforts in the field regarding the two methods, see \cite{2021ARA&A..59..117R}. Of the currently operating X-ray missions, NuSTAR (\citealt{2013ApJ...770..103H}) is the most ideally suited for measuring the features of relativistic reflection, namely the broadened Fe K$\alpha$ line, present at 6.4 keV for neutral gas and at progressively higher energies up to 6.97 keV for Fe XXVI, and the Compton hump, a broad energy excess above $\sim20$ keV. 

XTE J2012+381 was first discovered in 1998 using the RXTE All-Sky Monitor by \cite{1998IAUC.6920....1R}, with a candidate optical counterpart being quickly identified, but later confirmed by \cite{1999MNRAS.305L..49H}, which classified the outburst as soft X-ray transient. \cite{1998IAUC.6927....2W} analyzed an ASCA observation of the source and obtained a good fit to the spectrum using a disk blackbody and a power-law component model, and claimed it to be a black hole candidate. Later, \cite{2000A&A...362L..53V} analyzed 24 RXTE observations of XTE J2012+381 obtained throughout the 1998 outburst and claimed the presence of excess broadened emission around 6.4 keV. Based on the spectral features measured from five BeppoSAX observations, \cite{2002A&A...384..163C} placed a lower limit on the mass of the BH in XTE J2012+381 of $M \gtrsim22d_{10}\;M_\odot$ for a maximally spinning BH, where $d_{10}$ is the distance to the system in units of 10kpc. The Gaia (\citealt{2016A&A...595A...1G}) measurement of the parallax of XTE J2012+381 is $p=0.1859\pm0.0719\;\rm mas$, equivalent to a distance to the system of $d=5.4\pm2.1\;\rm kpc$.  Given the Gaia distance estimate and the estimate of \cite{2002A&A...384..163C}, we can place a lower limit on the BH mass of $M \gtrsim11.8\pm4.6M_\odot$.

XTE J2012+381 entered an outburst phase again in late 2022. This outburst was first detected by the MAXI/GSC nova alert system on December 25, 2022 (\citealt{2022ATel15826....1K}), and confirmed using the Swift XRT instrument on December 26, 2022 (\citealt{2022ATel15827....1K}). We obtained two NuSTAR observations of XTE J2012+381, and NICER (\citealt{2016SPIE.9905E..1HG}) monitored the source throughout the outburst. Motivated by the previous reports of the presence of relativistic reflection features in the spectra of this source during the previous outburst, we attempted to use the relativistic reflection method to measure the spin of the BH candidate XTE J2012+381. The summary of observations used in the analysis is presented in Section \ref{sec:obs} and our analysis methods and results are presented in section \ref{sec:analysis}. In Section \ref{sec:disc} we discuss the implications of this result on the broader stellar mass BH spin distribution.
 
\section{Observations and Data Reduction} \label{sec:obs}
We observed XTE J2012+381 twice using NuSTAR, obtaining an exposure on December 29, 2022 under ObsID 80802344002 and an exposure on January 18 2023 under ObsID 80802344004. We analyzed the NuSTAR observations using the routines in HEASOFT v6.29c through the NUSTARDAS pipeline v2.1.1 and CALDB v20211103. We extracted the source spectra from circular regions centered on the source position with radii of 120”, and we used regions of the same size for extraction of background rates. We grouped the spectra using the “ftgrouppha” ftool, through the optimal binning scheme described by \cite{2016A&A...587A.151K}. We continued analyzing the NuSTAR spectra in the 3-70 keV and 3-60 keV bands, respectively, as the spectra obtained during the two observations were background-dominated at higher energies. We chose to analyze the NuSTAR observations using these versions of the calibration software in order to maintain consistency with the larger sample presented in \cite{2023ApJ...946...19D}. However, we note that using spectra extracted using the latest calibration software available produces fully consistent results.

NICER tracked the evolution of the outburst by taking 105 observations of the source over the first 155 days between the first detection and May 29th. We analyzed the observations using the NICERDAS v10 pipeline in HEASOFT v6.31 and CALDB xti20221001. We ran the \texttt{nicerl2} pipeline by excluding the detectors 14 and 34.  During many observations, the NICER detectors were dominated by optical loading at low energies, producing residuals that cannot be accounted for using physical models under 1keV, regardless of the limit placed on the ``undershoots" in the NICER detector. Therefore, as the spectrum below 1keV cannot be properly constrained due to optical loading, we constrained the allowed ``undershoot" rates to be as high as 500 in order to not sacrifice the quality of the data at energies above 1keV, and only fit the NICER spectra down to 1keV. We set the allowed ``overshoot" rates to be as high as 1.5. We then extracted the source spectra and the associated RMF and ARF files using the \texttt{nicerl3} pipeline, and we accounted for background emission using the SCORPEON model. We fit the spectra in the 1-10 keV band. 

\section{Analysis and Results} \label{sec:analysis}
We ran the spectral analysis in XSPEC v12.12.0g (\citealt{1996ASPC..101...17A}) by minimizing the $\chi^2$ statistic. We independently fit the spectra obtained from all the NICER observations and the two NuSTAR Focal Plane Modules (FPM) from the two observations of the source. The initial model that we used describes an absorbed disk black body plus a power law component, \texttt{TBabs*(diskbb+powerlaw)}. This model includes the multiplicative component \texttt{TBabs} (\citealt{2006HEAD....9.1360W}) to account for the interstellar absorption using abundances computed by \cite{2000ApJ...542..914W} and photoionization cross sections computed by \cite{1996ApJ...465..487V}. 

\begin{figure*}[ht!]
    \centering
    \includegraphics[width= 0.8\textwidth]{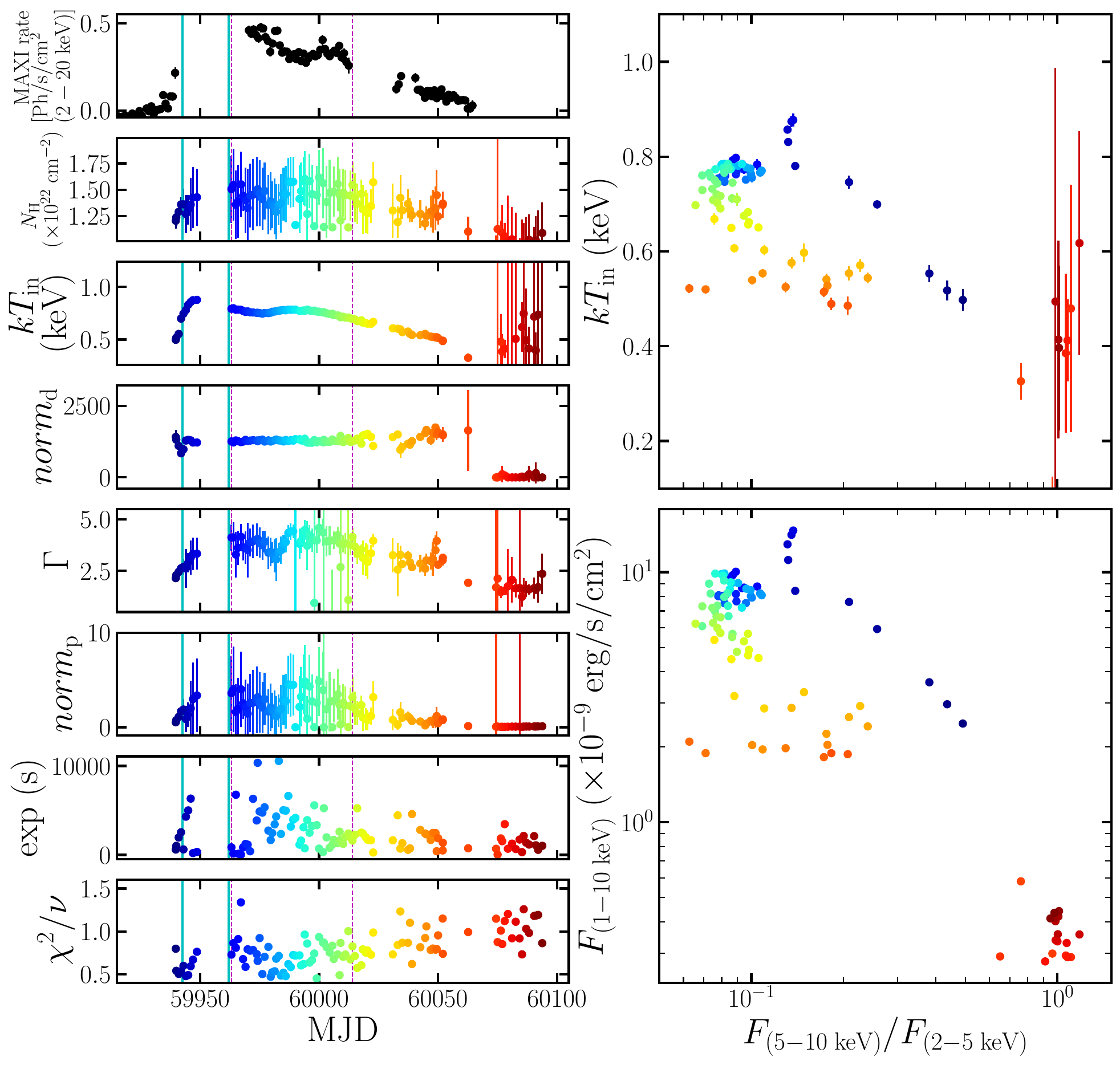}
    \caption{Left: The MAXI light curve in the 2-20 keV band of the 2022-2023 outburst of XTE J2012+381 (top). The following five panels show the evolution of the measured Galactic column density $N_{\rm H}$, the inner disk temperature and normalization of the \texttt{diskbb} component, the power-law index $\Gamma$, the normalization of the \texttt{powerlaw} component, obtained when fitting the NICER spectra of the source with the model \texttt{TBabs*(diskbb+powerlaw)}. The seventh panel shows the exposure of the NICER observations analyzed, and the eighth panel shows the reduced $\chi^2$ returned by the fits to the NICER spectra. The colors of the points represent time evolution. The vertical cyan lines show the dates of the two NuSTAR observations of XTE J2012+381 analyzed in this paper. The observations between the dashed vertical magenta lines were combined to produce the spectrum shown in Figure \ref{fig:combined_residuals}. Right: The evolution of the measured disk temperature of the source (top) and of the 1-10 keV flux (bottom) vs. the hardness defined as the ratio of the 5-10 keV flux to the 2-5 keV flux. In the top panel, we omitted the measurements which had an uncertainty larger than 0.5 keV. The colors of the points are the same as in the left panels and represent the time evolution of the source.}
    \label{fig:nicer_lc}
\end{figure*}

The top left panel in Figure \ref{fig:nicer_lc} shows the MAXI light curve of the outburst of XTE J2012+381, in the 2-20 keV band. The two cyan vertical lines represent the dates of the two NuSTAR observations of the source. The following panels on the left show the time evolution of the measurements of the Galactic column density, the accretion disk temperature and normalization, and of the power law index and normalization in the fits to the NICER spectra. The last two panels on the left in Figure \ref{fig:nicer_lc} show the effective exposure of the NICER observations analyzed, and the reduced $\chi^2$ produced when fitting the NICER spectra. The right panels show the link between the evolution of the measured temperature of the \texttt{diskbb} component (top) and the 1-10 keV flux (bottom) as a function of the hardness ratio, computed as the ratio of the fluxes in the 5-10 keV band and the 2-5 keV band. The colors of the points track the time evolution, similarly to the panels on the left. The outburst begins in an already relatively soft state, but evolves similarly to other BH outbursts, following a ``Q" shape in this plot. However, \cite{2023ATel15847....1R} reported an INTEGRAL detection of XTE J2012+381 on December 23rd, 2022 (3 days before the first NICER observation). During this X-ray observation, the source was in a harder state and well detected up to high (150 keV) energies, suggesting that the source transitioned from a hard to soft state in the very early stages of its outburst.

When fitting NuSTAR spectra, it is often customary to allow the presence of a normalization constant to account for the difference between the spectra from the two detectors. However, we did not include a \texttt{constant} component in our models and instead we allowed the normalizations of the \texttt{diskbb} and the \texttt{powerlaw} components to vary independently. This introduces an additional free parameter when compared to adding a \texttt{constant} component to the model, but the quality of the fits is often superior to simply allowing a constant offset between the spectra. When allowing the normalizations of the components to vary independently, they generally take values within a few percent of each other.  The residuals produced when fitting the NuSTAR spectra show clear signs of relativistic reflection.

To account for the relativistic reflection features, we replaced the \texttt{powerlaw} component in our baseline model with different flavors of the \texttt{relxill} v.1.4.3 family of models (\citealt{2014MNRAS.444L.100D, 2014ApJ...782...76G}). A complete description of the models can be found on the \texttt{relxill} website\footnote{\url{http://www.sternwarte.uni-erlangen.de/~dauser/research/relxill/}}, Section 3.1 in \cite{2021ApJ...920...88D}, or Appendix A in \cite{2023ApJ...947...39D}. While newer versions of the \texttt{relxill} models include the effect of returning radiation, works such as \cite{2022MNRAS.514.3965D} and \cite{2023arXiv230312581R} concluded that the measured spin of the compact object in the system is unaffected by the inclusion of returning radiation. Therefore, in order to ensure consistency of our analysis with the pipeline of \cite{2023ApJ...946...19D}, we chose to use the same version of \texttt{relxill}. Similarly, for consistency with the large-scale analysis of \cite{2023ApJ...946...19D}, we initially explored the effects of replacing the \texttt{powerlaw} component in our initial fits with six different flavors of the \texttt{relxill} family of models: \texttt{relxill}, \texttt{relxillCp}, \texttt{relxilllp}, and the \texttt{relxillD} version with the accretion disk density fixed to $n=10^{15}$, $10^{17}$, and $10^{19}\;\rm cm^{-3}$.

Given the existing mass and distance estimates (presented in Section \ref{sec:intro}) and the inferred fluxes based on the two NuSTAR observations, the source falls within the range of luminosity for which based on theoretical, numerical, and observational results (see, e.g., \citealt{2008ApJ...675.1048R, 2013MNRAS.431.3510S, 2015ApJ...813...84G, 2016ApJ...819...48S}) it is expected that the inner disk radius extends near to the innermost stable circular orbit (ISCO) of the BH: $10^{-3}\lesssim L/L_{\rm Edd}\lesssim 0.3$, where $L_{\rm Edd}$ represents the Eddington luminosity, and $L$ represents the luminosity of the source. Therefore, throughout our spectral analysis, we set the inner disk radius to be that of the ISCO, $r_{\rm in}=r_{\rm ISCO}$. We fixed the outer disk radius at $r_{\rm out}=990\;r_g$. We allowed all other parameters in the models to vary freely. 

We fit the NuSTAR spectra from the two observations both independently, and jointly with NICER observations of the source taken closely in time to the NuSTAR exposures. We fit NuSTAR obsID 80802344002 together with NICER obsID 5203600104, which overlapped with the NuSTAR observation, and NuSTAR obsID 80802344004 together with NICER obsID 5203600114, which was taken three days after the second NuSTAR observation. We chose this NICER observation over other, closer in time to the second NuSTAR observation, as it had a significantly longer exposure. We applied the array of six \texttt{relxill} flavors to both the NuSTAR spectra alone, and to the NICER and NuSTAR spectra together.

The spectra from the first NuSTAR observation (80802344002) are dominated by a high-energy component. Fitting the NuSTAR spectra jointly with the NICER spectrum from obsID 5203600104 with the six variants of the reflection model produces good fits. The best-performing model was \texttt{TBabs*(diskbb+relxill)} producing $\chi^2/\nu=549.15/556=0.99$, followed closely by the \texttt{relxillD} variant with $\log(n)=19$ producing $\chi^2/\nu=550.09/557=0.99$ and the \texttt{relxillD} variant with $\log(n)=15$, with $\chi^2/\nu=569.18/557=1.02$. Given the low-energy coverage provided by the addition of NICER spectra, we also tested the effects of modeling the accretion disk with a more physically accurate component, by replacing the \texttt{diskbb} component in the best performing model with the \texttt{kerrbb} model (\citealt{2005ApJS..157..335L}). In the \texttt{kerrbb} component, we fixed the BH mass to 11.8$M_\odot$, the distance to the BH to 5.2 kpc, and linked the BH spin and inner disk inclination between the \texttt{kerrbb} and \texttt{relxill} components. This returned an improved statistic of $\chi^2/\nu=537.3/555=0.97$ for the \texttt{TBabs*(kerrbb+relxill)} model. The other models tested performed worse in terms of statistic, but produced relatively similar parameter constraints.

When using the six models to fit the NuSTAR spectra from the first observation alone, the best-performing model was \texttt{TBabs*(diskbb+relxill)}, producing $\chi^2/\nu=452.45/420=1.08$, followed closely by the \texttt{relxillD} variant with $\log(n)=15$ producing $\chi^2/\nu=459.1/421=1.09$ and by the \texttt{relxillCp} variant with $\chi^2/\nu=460.32/420=1.10$. The other three models tested performed worse. Despite the lack of low-energy coverage under 3 keV when not including NICER spectra to the fit, we tested the effects of replacing the \texttt{diskbb} component with the \texttt{kerrbb} one. This returned $\chi^2/\nu=452.12/419=1.08$, which formally improves $\chi^2$, but due to the extra free parameter, the improvement over the the model assuming a simplistic disk treatment is not statistically significant.

Fitting the second pair of NuSTAR and NICER spectra with the six model variants again produces reasonable fits. By far, the best-performing model is \texttt{TBabs*(diskbb+relxill)}, which returns $\chi^2/\nu=482.14/459=1.05$. As the observations occurred while the source was in a disk-dominated state, one would naively expect that the improvement of replacing the \texttt{diskbb} component with \texttt{kerrbb} would be more significant in this case. However, fitting the spectra with the \texttt{TBabs*(kerrbb+relxill)} model produces a worse fit, returning $\chi^2/\nu=493.15/458=1.08$. This result is surprising, given how the \texttt{kerrbb} component is more complex, with more free parameters, and one would expect the fit to converge to at least the same value of $\chi^2$. However, the increased complexity of the component originating from multiple parameters which are strongly correlated to produce similar spectral features, paired with the limited data quality makes the parameter space difficult to explore and the fit prone to converging to local $\chi^2$ minima, as opposed to the global best-fit solution. Furthermore, it is important to note that at this point, the majority of the contribution to $\chi^2$ comes from instrumental residuals in the NICER spectrum and from possible differences between the FPMA and FPMB instruments on NuSTAR. Nevertheless, the reflection component remains similar regardless of the assumption of disk model.

\begin{figure*}[ht!]
    \centering
    \includegraphics[width= 0.71\textwidth]{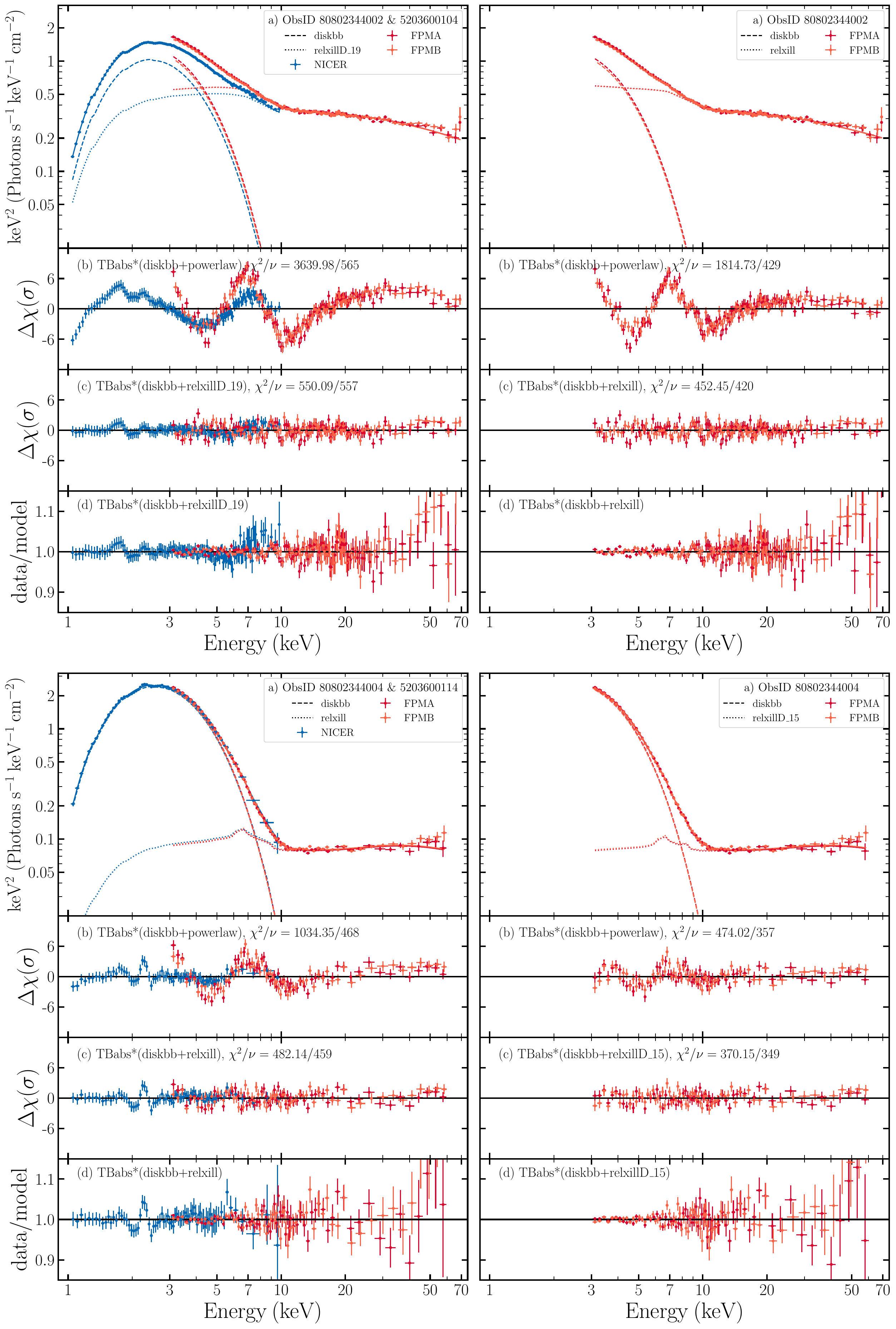}
    \caption{Sub-panels (a) shows the unfolded spectra of XTE J2012+381. The blue points represent the NICER spectra, and the different shades of red indicate the spectra from the NuSTAR FPMA and FPMB detectors. The solid lines represent the total best-fit models, while the dashed and dotted lines show the contributions to the model by the \texttt{diskbb} and \texttt{relxill/relxillD} components, respectively.  Sub-panels (b) shows the residuals in terms of $\sigma$ for the \texttt{TBabs*(diskbb+powerlaw)} model. The residuals show clear indication of relativistic reflection for both observations. Sub-panels (c) shows the residuals of the best-fit models, \texttt{TBabs*(diskbb+relxill)} or \texttt{TBabs*(diskbb+relxillD)}, respectively, for each observation, together with the statistic produced by the models. Sub-panels (d) show the ratio of data to model for the best-fit models. The left panels show the spectra and residuals when fitting NICER and NuSTAR observations jointly, while the right panels show the results obtained when fitting the NuSTAR observations alone. The top panels indicate the first epoch (December 29, 2022), while the bottom panels show the second epoch (January 18 and 21, 2023 for NuSTAR and NICER respectively).}
    \label{fig:delchi_all}
\end{figure*}

When fitting the NuSTAR spectra of the second observation without including the NICER spectrum, the fits from multiple models converge to the same solution, in the same region of the parameter space. The models \texttt{TBabs*(diskbb+relxillD)} with $\log(n)=15$, \texttt{TBabs*(diskbb+relxillD)} with $\log(n)=19$, \texttt{TBabs*(diskbb+relxill)}, and \texttt{TBabs*(kerrbb+relxill)} produce $\chi^2/\nu=370.15/349=1.06$, $\chi^2/\nu=371.85/349=1.07$, $\chi^2/\nu=372.62/348=1.07$, and $\chi^2/\nu=372.27/346=1.08$, respectively. Particularly peculiar about these results is that the \texttt{relxillD} with $\log(n)=15$ variant performs better than the \texttt{relxill} variant, since the only difference between the two is that \texttt{relxill} allows the high-energy cutoff of the incident power-law spectrum to vary, while \texttt{relxillD} fixes it at 300 keV, while both these particular variants have fixed $\log(n)=15$. This suggests that the fit using the \texttt{TBabs*(diskbb+relxill)} was indeed stuck in a local minimum, but with a fit statistic very similar to the global minimum. Similarly to the case of the first observation, when lacking low-energy coverage, replacing the \texttt{diskbb} component with the \texttt{kerrbb} one does not significantly influence the quality of the fit or the reflection parameter combination. 

Following the pipeline described in \cite{2023ApJ...946...19D}, we ran a Markov Chain Monte Carlo (MCMC) analysis of the parameter space on the best fits produced for each observation. For specifics regarding the MCMC analysis, please refer to Section 2.2 in \cite{2023ApJ...946...19D}. We ran the MCMC analysis on the 3 best-performing models that describe the thermal emission from the accretion disk using the \texttt{diskbb} component and on the model that describes the disk emission using \texttt{kerrbb} and the coronal and reflected emission using \texttt{relxill}. We computed the Deviance Information Criterion (DIC; \citealt{DIC_text}) based on all MCMC runs, and we use this number to quantify the goodness of fit and to distinguish between models that perform similarly in terms of statistic produced. 

When fitting the first NuSTAR observation alone, the \texttt{TBabs*(diskbb+relxill)} model produces DIC=479.51, the \texttt{TBabs*(diskbb+relxillD)} variant with $\log(n)=15$ produces DIC=486.84, the \texttt{TBabs*(diskbb+relxillCp)} model produces DIC=491.47, and the \texttt{TBabs*(kerrbb+relxill)} model produces DIC=576.53. When fitting this NuSTAR observation jointly with the NICER observation 5203600104, the \texttt{TBabs*(diskbb+relxillD)} variant with $\log(n)=19$ produces DIC=579.53, the \texttt{TBabs*(diskbb+relxill)} model produces DIC=582.43, the \texttt{TBabs*(kerrbb+relxill)} model produces DIC=583.5, and the \texttt{TBabs*(diskbb+relxillD)} variant with $\log(n)=15$ produces DIC=598.91.

When fitting the second NuSTAR observation alone, the \texttt{TBabs*(diskbb+relxillD)} variant with $\log(n)=15$ produces DIC=396.45, the \texttt{TBabs*(diskbb+relxillD)} variant with $\log(n)=19$ produces DIC=400.47, the \texttt{TBabs*(diskbb+relxill)} model produces DIC=401.58, and the \texttt{TBabs*(kerrbb+relxill)} model produces DIC=545.25. When including the NICER observation 5203600114 and fitting the spectra jointly, the \texttt{TBabs*(diskbb+relxill)} model returns DIC=518.88, the \texttt{TBabs*(diskbb+relxillD)} variant with $\log(n)=19$ produces DIC=519.99, the \texttt{TBabs*(diskbb+relxillD)} variant with $\log(n)=15$ produces DIC=521.36, and the \texttt{TBabs*(kerrbb+relxill)} model produces DIC=529.23. 

The top sub-panels in Figure \ref{fig:delchi_all} show the unfolded spectra of the observations taken during the two epochs, which we analyzed in this paper. The right panels show only the NuSTAR FPMA and FPMB spectra through the red points, while the left panels include the NICER spectra, shown through the blue points. The solid lines represent the total best-fit models, the dashed lines represent the contribution of the \texttt{diskbb} component in the models, and the dotted lines represent the contributions of the best-performing reflection component. Sub-panels b) and c) show the contribution to the residuals in terms of sigma when fitting the spectra using the \texttt{TBabs*(diskbb+powerlaw)} model, which does not account for relativistic reflection, and with the best-performing models which does account for reflection, respectively, together with the fit statistic produced. Sub-panels d) show the ratio of data to model for the best-performing reflection models. 

Visually, the highest contributions to the residuals come from unaccounted instrumental features in the NICER spectra, around the Al edge at 1.56 keV and the Si edge at 1.84 keV, due to FPM detector features, and around the Au M edge around 2.2 keV due to the reflectivity of gold M shells in the NICER X-ray Concentrator (XRC) optics. However, these residuals have a relatively low impact on the total statistic of the fit. We tested the effect of trying to account for those residuals by adding \texttt{gaussian} components to the best-fit models. For the first observation, the fit prefers the addition of a \texttt{gaussian} component at 2.27 keV for the Au M edge at 2.2 keV, and another \texttt{gaussian} component at 1.7 keV, at the average of the 1.56 keV Al edge and the 1.84 Si edge, accounting for both features. The addition of the two components improves the quality of the fit by $\Delta \chi^2=16$ for 6 additional free parameters. For the second observation, adding a \texttt{gaussian} at 2.26 keV improves the fit by $\Delta \chi^2=13$ for three extra free parameters. We note that all the improvement comes for the NICER spectrum, and the fit to the NuSTAR spectra return the same $\chi^2$, suggesting that the continuum is constrained in the same way regardless of the correction for instrumental features. As the reflection parameters are nearly entirely determined by features that fall above 3 keV and seeing how the underlying continuum is constrained the same way regardless of the addition of correction \texttt{gaussian} components, we chose to continue our analysis without the extra components. This choice was made in order to reduce the complexity of the models, and make variations in information criteria be more impacted by the ability to constrain reflection features as opposed to instrumental features.

\begin{figure*}[ht!]
    \centering
    \includegraphics[width= 0.95\textwidth]{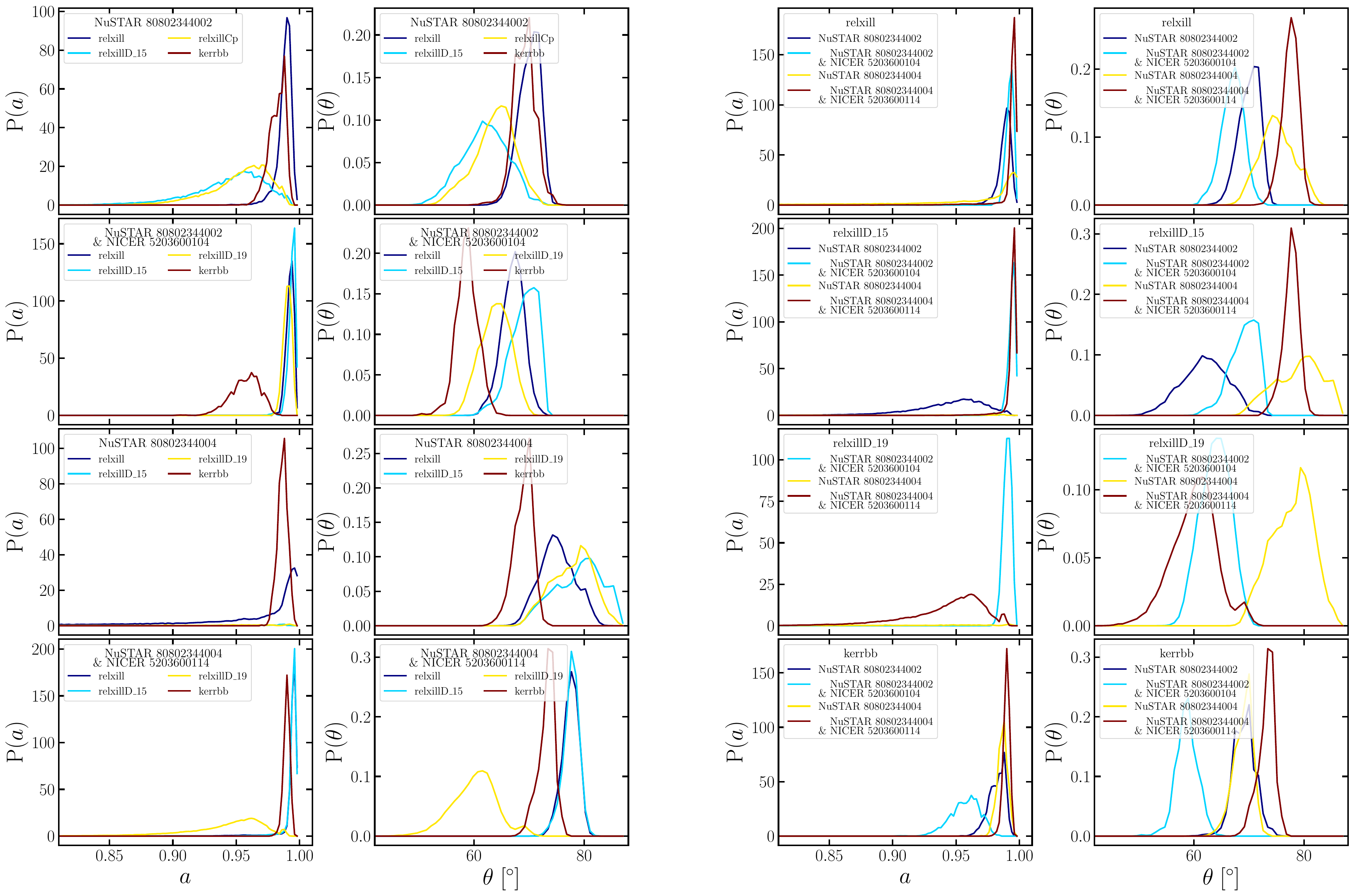}
    \caption{Histograms of the posterior distributions in the MCMC analysis for spin and inclination. The two panels on the left show the distributions grouped by observation analyzed, when using different models, with the first and third rows indicating fits to the NuSTAR observations alone, and the second and fourth rows indicating joint NuSTAR and NICER fits. The two panels on the right show the distributions grouped by the model used to fit the observations.}
    \label{fig:posteriors_all}
\end{figure*}

Figure \ref{fig:posteriors_all} shows the 1-dimensional histograms of the posterior distributions for the spin (left sub-panels) and inclination (right sub-panels) based on the MCMC runs. The left panels of Figure \ref{fig:posteriors_all} show the posterior distributions for the two epochs, when fitting only the NuSTAR observations (first and third sub-panels, in the downward direction) and both the NuSTAR and NICER observations (second and fourth sub-panels, in the downward direction). These are grouped by observation analyzed. The right panels in Figure \ref{fig:posteriors_all} show the posterior distributions produced by the different model variations analyzed when treating the observations from the two different epochs, when analyzing the NuSTAR observations alone, and when also fitting the NICER spectra jointly with the NuSTAR ones. These are grouped by the model used.

\begin{figure*}[ht!]
    \centering
    \includegraphics[width= 0.8\textwidth]{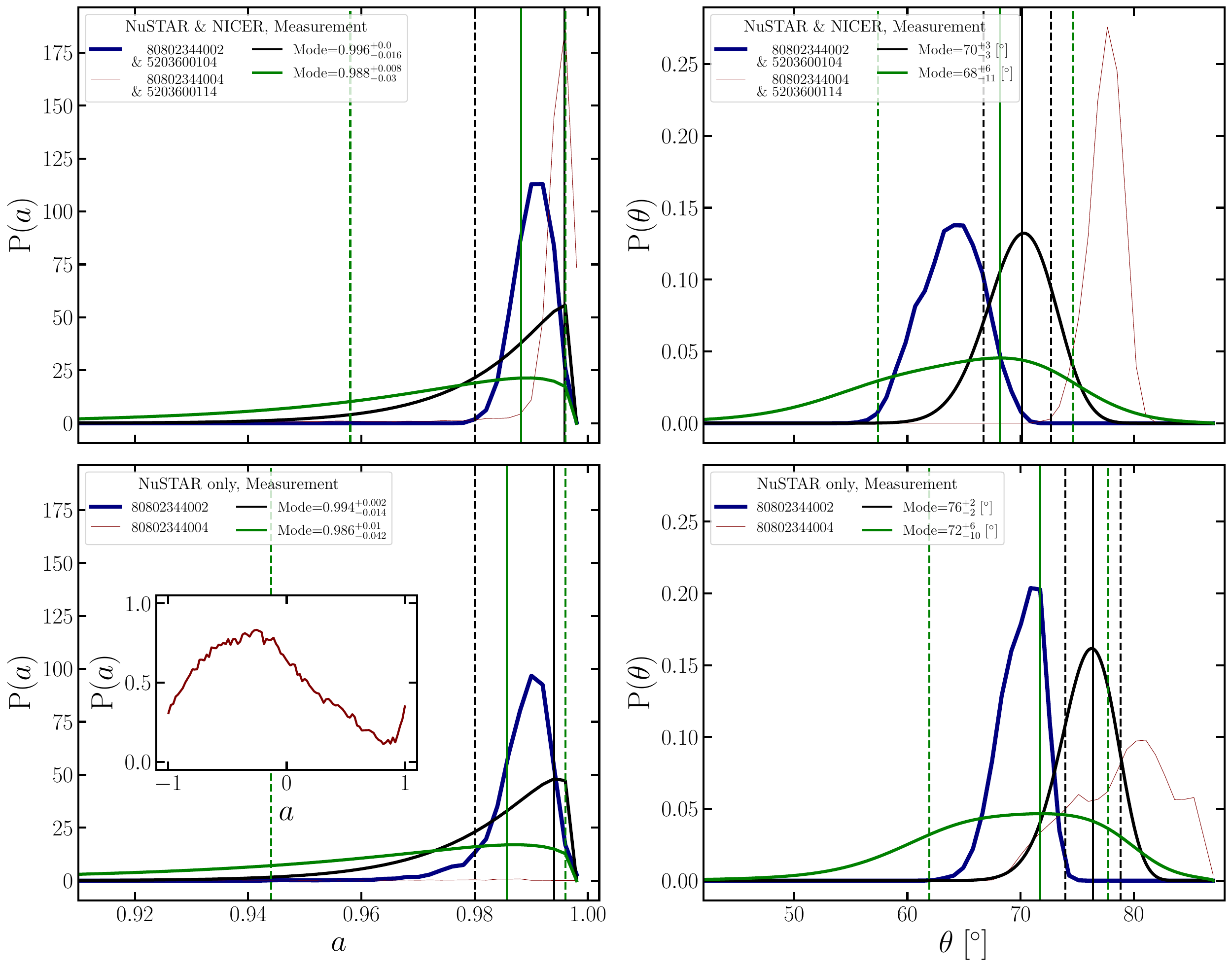}
    \caption{The left panels shows the posterior distributions resulting from the MCMC analysis of XTE J2012+381 for spin, while the right panels shows the posterior distributions for the inclination of the inner accretion disk in the model. The width of the lines is proportional to the ratio of the reflected flux to the total flux in the 3-79~keV band, which were used as weighting when combining the posterior distributions. The top panels indicate the results obtained from joint NICER and NuSTAR fits, while the bottom panels represent the results obtained from NuSTAR-only fits. The blue and red lines represent the posteriors obtained when fitting the two sets of observations analysed, and the black curves represent the combined inferred distributions, derived as highlighted by \cite{2023ApJ...946...19D}. The green lines represent the combined distributions obtained as highlighted in this paper, which better account for systematic variations between the results obtained from the independent observations. The solid vertical black and green lines represent the modes of the combined distributions, and the dashed vertical black and green lines represent the $1\sigma$ credible intervals of the respective measurements. The insert in the lower left panel shows the complete posterior distribution for the spin obtained when fitting the second NuSTAR observation alone with the \texttt{TBabs*(diskbb+relxillD)} model, which obtains two solutions similar in terms of statistic produced, but with very different spin constraints. Nevertheless, as the first observation (blue) weights significantly more in the combining algorithm due to stronger reflection, the combined distribution still significantly favors a high spin, but with a broader lower limit on the credible interval.}
    \label{fig:combined}
\end{figure*}

Similarly to the prescription of \cite{2023ApJ...946...19D}, we combined the posterior distributions of the best-performing models in terms of DIC for the BH spin and inner disk inclination for the two epochs. The top panels in Figure \ref{fig:combined} show the histograms of the posterior samples for the spin (left) and inclination (right) when fitting the NuSTAR and NICER observations jointly, with the blue curves indicating the measurements based on observations taken during the first epoch and the red curves indicating the measurements based on observations taken during the second epoch. The bottom panels show the posterior distributions obtained when fitting only the NuSTAR spectra from the two epochs. The line width of the blue and red curves indicate the weighting used when combining the measurements with a beta distribution, which was calculated to be proportional to the ratio of the reflected to total flux in the 3-79 keV band. The black curves in Figure \ref{fig:combined} indicate the combined beta distribution obtained based on the mode of the posterior distributions of the parameters $a$ and $b$ describing the beta distribution, according to the method used in \cite{2023ApJ...946...19D}. The vertical solid and dashed lines represent the mode and the $\pm 1\sigma$ credible intervals of the combined spin and inclination distribution. 

Furthermore, to better encapsulate the differences between the measurements produced by the observations of the source during the two epochs owing to systematic uncertainties, we also combined the measurements using a novel method. Upon running the Bayesian algorithm used to combine the individual measurements into a single beta distribution (the black curve), we randomly selected 10000 of the posterior samples generated while running the algorithm and averaged them. These resulting averaged beta distributions are shown in Figure \ref{fig:combined} through the solid green curves, and the modes and $\pm 1\sigma$ credible intervals are shown through the green vertical solid and dashed lines.

The insert in the bottom left panel of Figure \ref{fig:combined} shows the complete histogram of the posterior samples for the BH spin when fitting the second NuSTAR observation alone with the \texttt{relxillD} model with with $\log(n)=15$, which performs best in terms of DIC among the models used to fit this spectrum. While running the MCMC analysis, the walkers discovered a second combination of parameters that is similarly favored in terms of statistic produced to the one that was used to initialize the walkers. While the initial best fit favored a high BH spin, the second solution for this model favors a moderately low spin, consistent with a non-rotating BH. While the high-spin solution also has a high inner emissivity index $q_1$, the low-spin solution takes low values of $q_1$. As suggested by \cite{2014MNRAS.439.2307F}, such solutions are to be treated as lower limits only, as for flat emissivity parameters, in order to match the flux, the inner disk radius in the models is pushed outward. When the inner disk radius is linked to the size of the ISCO, this translates to a lowering in measured BH spin. Despite the two solutions producing similar $\chi^2$ values, the walkers in the MCMC analysis favor the low-spin solution, as the parameter space is wider and easier to explore. The likelihood space for the high-spin solution is very narrow, making it easier for the walkers to leave the high-spin solution and explore the low-spin one, but very difficult to return to the high-spin region of the parameter space. This combination of parameters only produces a good fit for the \texttt{relxillD} variants when fitting the second NuSTAR observation alone. Low-spin solutions are strongly disfavored when fitting the same NuSTAR observation jointly with a NICER spectrum, when fitting the NuSTAR spectra alone with other \texttt{relxill} variants, or when fitting the other NuSTAR observation, either alone or jointly with a NICER observation, with any \texttt{relxill} variant. As the second has a strength of reflection much smaller compared to the first observation, despite the fact that many of the posterior samples in the MCMC run prefer a low spin, the combined measurement still yields a high value. However, the lower limit of the credible interval takes a lower value, suggesting that in this case, the BH spin is poorly constrained.

\begin{deluxetable*}{c|cccc|c}
\tablecaption{Results of the MCMC analysis}
\label{table:results}
\tablewidth{\textwidth} 
\tabletypesize{\scriptsize}
\tablehead{
\colhead{ObsID} & \colhead{80802344002 \& 5203600104} & \colhead{80802344002} & \colhead{80802344004 \& 5203600114} & \colhead{80802344004} & \colhead{80802344004}
}
\startdata\\
instrument & NuSTAR \& NICER & NuSTAR & NuSTAR \& NICER & NuSTAR & NuSTAR\\
\hline
model & relxillD-19 & relxill & relxill & relxillD-15 & relxill \\
\hline
$N_H\;[\times10^{22}\;\rm cm^{-2}]$ & $1.83_{-0.02}^{+0.02}$ & $0.8_{-0.2}^{+0.3}$ & $1.750_{-0.008}^{+0.010}$ & $0.7_{-0.1}^{+0.2}$ & $0.8_{-0.1}^{+0.2}$ \\
\hline
$kT_{\rm in}\;[\rm keV]$ & $0.726_{-0.004}^{+0.006}$ & $0.79_{-0.01}^{+0.02}$ & $0.789_{-0.001}^{+0.002}$ & $0.815_{-0.004}^{+0.005}$ & $0.814_{-0.006}^{+0.005}$ \\
$\rm norm_{\rm d,A} \; [\times10^2]$ & $10.6_{-0.4}^{+0.3}$ & $5.5_{-0.3}^{+1.2}$ & $13.8_{-0.2}^{+0.2}$ & $10.8_{-0.4}^{+0.4}$ & $10.8_{-0.5}^{+0.6}$ \\
\hline
$q_1$ & $7.0_{-0.7}^{+0.9}$ & $9.9_{-1.0}^{+0.1}$ & $9.9_{-1.1}^{+0.1}$ & $3.5_{-0.5}^{+2.7}$ & $3.4_{-0.4}^{+4.3}$ \\
$q_2$ & $2.4_{-1.2}^{+0.6}$ & $1.7_{-0.6}^{+0.3}$ & $1.93_{-0.60}^{+0.07}$ & $1.9_{-0.2}^{+0.2}$ & $1.9_{-0.3}^{+0.2}$ \\
$R_{\rm br} \;[r_g]$ & $92_{-50}^{+8}$ & $3.3_{-0.3}^{+2.2}$ & $2.9_{-0.6}^{+1.2}$ & $13_{-5}^{+7}$ & $2.6_{-0.6}^{+3.8}$ \\
$a$ & $0.990_{-0.003}^{+0.003}$ & $0.989_{-0.004}^{+0.003}$ & $0.994_{-0.002}^{+0.002}$ & $-0.2_{-0.6}^{+0.4}$ & $0.99_{-0.14}^{+0.01}$ \\
$\theta\;[^\circ]$ & $65_{-3}^{+2}$ & $71.2_{-2.9}^{+0.8}$ & $77_{-1}^{+2}$ & $80_{-6}^{+3}$ & $74_{-3}^{+4}$ \\
$\Gamma$ & $2.21_{-0.02}^{+0.01}$ & $2.28_{-0.02}^{+0.01}$ & $2.30_{-0.05}^{+0.03}$ & $2.06_{-0.03}^{+0.05}$ & $2.13_{-0.04}^{+0.07}$ \\
$\log(\xi)$ & $3.86_{-0.08}^{+0.09}$ & $4.21_{-0.16}^{+0.05}$ & $0.3_{-0.3}^{+0.6}$ & $\leq0.004$ & $0.3_{-0.3}^{+1.3}$ \\
$A_{\rm Fe}$ & $9.9_{-1.2}^{+0.1}$ & $9.6_{-3.1}^{+0.4}$ & $9.9_{-1.5}^{+0.1}$ & $9.8_{-2.3}^{+0.2}$ & $8_{-1}^{+2}$ \\
$\rm E_{\rm cut}\;[\rm keV] $ & $300^*$ & $980_{-200}^{+20}$ & $970_{-300}^{+30}$ & $300^*$ & $900_{-400}^{+100}$ \\
R & $1.3_{-0.2}^{+0.2}$ & $1.4_{-0.2}^{+0.3}$ & $3.8_{-0.7}^{+0.6}$ & $1.8_{-0.6}^{+0.6}$ & $1.5_{-0.3}^{+0.7}$ \\
$\rm norm_{\rm r,A} [\times10^{-3}]$ & $5.5_{-0.4}^{+0.5}$ & $6.0_{-0.5}^{+0.7}$ & $1.8_{-0.1}^{+0.1}$ & $1.23_{-0.04}^{+0.06}$ & $1.4_{-0.1}^{+0.1}$ \\
\hline
$\rm norm_{\rm d,B} [\times10^2]$ & $10.2_{-0.5}^{+0.3}$ & $5.2_{-0.3}^{+1.1}$ & $13.5_{-0.2}^{+0.2}$ & $10.7_{-0.4}^{+0.4}$ & $10.7_{-0.5}^{+0.5}$ \\
$\rm norm_{\rm r,B} [\times10^{-3}]$ & $5.4_{-0.4}^{+0.5}$ & $6.0_{-0.5}^{+0.7}$ & $1.8_{-0.1}^{+0.1}$  & $1.26_{-0.05}^{+0.06}$ & $1.4_{-0.1}^{+0.1}$ \\\\
\hline
$\rm norm_{\rm d,N} [\times10^2]$ &$8.5_{-0.3}^{+0.3}$  &\nodata  &$13.3_{-0.2}^{+0.1}$  &\nodata  &\nodata  \\
$\rm norm_{\rm r,N} [\times10^{-3}]$ &$4.8_{-0.3}^{+0.4}$  &\nodata  &$1.8_{-0.2}^{+0.2}$  &\nodata  &\nodata  \\
\hline
$\chi^2/\nu$ & $564_{-5}^{+5}(550.09)/557$ & $464_{-4}^{+6}(452.45)/420$ & $495_{-6}^{+7}(482.14)/459$ & $382_{-5}^{+5}(370.15)/349$ & $384_{-4}^{+6}(372.62)/348$ \\
\enddata
\tablecomments{In this table, we report the modes of the posterior distributions in the MCMC analysis, along with the $1\sigma$ credible intervals. For $\chi^2$, the number in parentheses indicates the best-fit $\chi^2$ value. For the normalization of the components, the subscripts d (from \texttt{diskbb}) and r (from \texttt{relxill}) represent the model that the values correspond to, while the subscripts A, B, and N stand for NuSTAR FPMA, NuSTAR FPMB, and NICER, respectively. For the \texttt{relxillD} variants of the models, the high-energy cutoff is set to 300 keV by construction, and we indicate this in the table with *.}
\end{deluxetable*}

Table \ref{table:results} shows the modes of the posterior distributions along with the $\pm 1\sigma$ credible intervals of the posterior distributions based on the MCMC analysis of the best-performing models for the two epochs, when including a NICER observation in the fit and when fitting the NuSTAR spectra alone. As shown in the insert in the bottom left panel in Figure \ref{fig:combined}, the preferred solution has a low spin. For comparison, we include the results produced when fitting only the spectra from the second NuSTAR observation with the default \texttt{relxill} flavor. In Appendix \ref{sec:corner} we show the 1D and 2D parameter space of spin, inclination, and $\chi^2$ based on the MCMC analysis of the joint NuSTAR-NICER fits, and also the complete corner plots of the MCMC runs that produced the best DIC values for both the joint NuSTAR-NICER fits and the NuSTAR only fits.

The measured Fe abundance is high, $A_{\rm Fe}\gtrsim7$, consistent in both observations, regardless of the inclusion low-energy coverage through NICER spectra in the analysis. One possible explanation for the enhanced Fe abundance is levitation of Fe ions by radiation pressure in the innermost regions of the accretion disk, which would enlarge the abundance of iron in the disk photosphere (\citealt{2012ApJ...755...88R}). The high energy cutoff is high and poorly constrained in both observations. The disk component and the Galactic absorption are constrained differently when including the soft NICER coverage to the fits. However, regardless of how the underlying disk continuum is accounted for, the relativistic reflection features produce similar spin constraints. This is further suggested by agreeing results being produced when replacing the \texttt{diskbb} component in the models with \texttt{kerrbb} - see Figure \ref{fig:posteriors_all}. While models that describe the contribution of the accretion disk through the \texttt{kerrbb} component are disfavored in terms of DIC due to the complexity of the model, the spin measurements agree well with those from models that describe the disk contribution through the simpler \texttt{diskbb} component.

The measured ionization parameter is high during the first observation, $\log(\xi)\sim4$, but low during the second observation, $\log(\xi)\lesssim1$. Fixing one ionization measurement in the model used to fit the other observation produces bad fits, suggesting that given the size of the parameter space of the models, the measured ionization values are required by the data. It is important to acknowledge that the change in measured ionization between the two epochs is likely not physical, as the ionizing flux did not change significantly between the two sets of observations, and it is unlikely that the accretion disk density changed by many orders of magnitude over such a short timescale. The more likely explanation for the combination of peculiar change in ionization parameter and elevated Fe abundance has to do with the increased reflection fraction during the second epoch, when the ionization is lower, but more importantly the relatively low density of the accretion disk material. While throughout our analysis we probed values of the disk density of $10^{15}-10^{19}\; \rm cm^{-3}$, it is likely that much higher disk densities would, in fact, help reconcile the apparently abnormal measurements of $\log(\xi)$ and $A_{\rm Fe}$ (see, e.g. \citealt{2018ApJ...855....3T}). 

To probe that, we attempted fitting the data with the \texttt{reflionx$\_$HD} model\footnote{The table models can be downloaded from \url{https://ftp.ast.cam.ac.uk/pub/mlparker/reflionx/}. We used the \texttt{reflionx\_HD\_nthcomp\_v2.fits} file.} (\citealt{2020MNRAS.498.3888J, 2021ApJ...909..146C}), which allows accretion disk densities up to $10^{22}\;\rm cm^{-3}$. In full \texttt{xspec} parlance, the model used was \texttt{TBabs*(diskbb+nthcomp+ relconv*atable\{reflionx$\_$HD\_nthcomp\_v2.fits\})}. This model does fit the data well, formally improving the value of $\chi^2$, but with most of the improvement coming at low energies, in the NICER band, and the quality of the fit being essentially unchanged for the two NuSTAR spectra. In this case, when fixing the accretion disk density to $10^{22}\;\rm cm^{-3}$, both observations produce consistent ionization measurements $\log(\xi)\sim3.5$ and reduced values of $A_{\rm Fe}\sim1.5$. However, as the model does not include a parameter than quantifies the reflection fraction, a direct comparison between the outputs of the two models is not trivial. Nevertheless, the spin measurement (coming from the \texttt{relconv} model, not the \texttt{reflionx$\_$HD} one) is again high, consistent with the measurement derived through our analysis. In the future, large scale studies and comparisons between models using different accretion disk densities will enable quantifying the systematic effect of the assumption regarding disk density and the measured BH spins. For now, as the \texttt{reflionx$\_$HD} model produces values consistent with our analyisis using the \texttt{relxill} family of models, and in the interest of maintaining consistency with the rest of the sample of \citealt{2023ApJ...946...19D}, we report the results of our analysis using \texttt{relxill} and defer the comparison with other families of models for future work.

Low and intermediate values of the ionization parameter lead to a narrower Fe K line profile (\citealt{1993MNRAS.262..179M, 2000PASP..112.1145F}). In contrast, increasing the inclination would broaden the blue wing of the line, while increasing the spin would broaden the red wing of the line. Furthermore, high inclination systems often show evidence of disk winds which produce absorption features around 7 keV (see, e.g., \citealt{2006ApJ...646..394M, 2012ApJ...746L..20K, 2012MNRAS.422L..11P, 2014ApJ...784L...2K, 2020ApJ...900...78D}). If present and unaccounted for, such an absorption feature could possibly lead to biased characterizations of the ionization, BH spin, and viewing inclination. 

\begin{figure}[ht!]
    \centering
    \includegraphics[width= 0.45\textwidth]{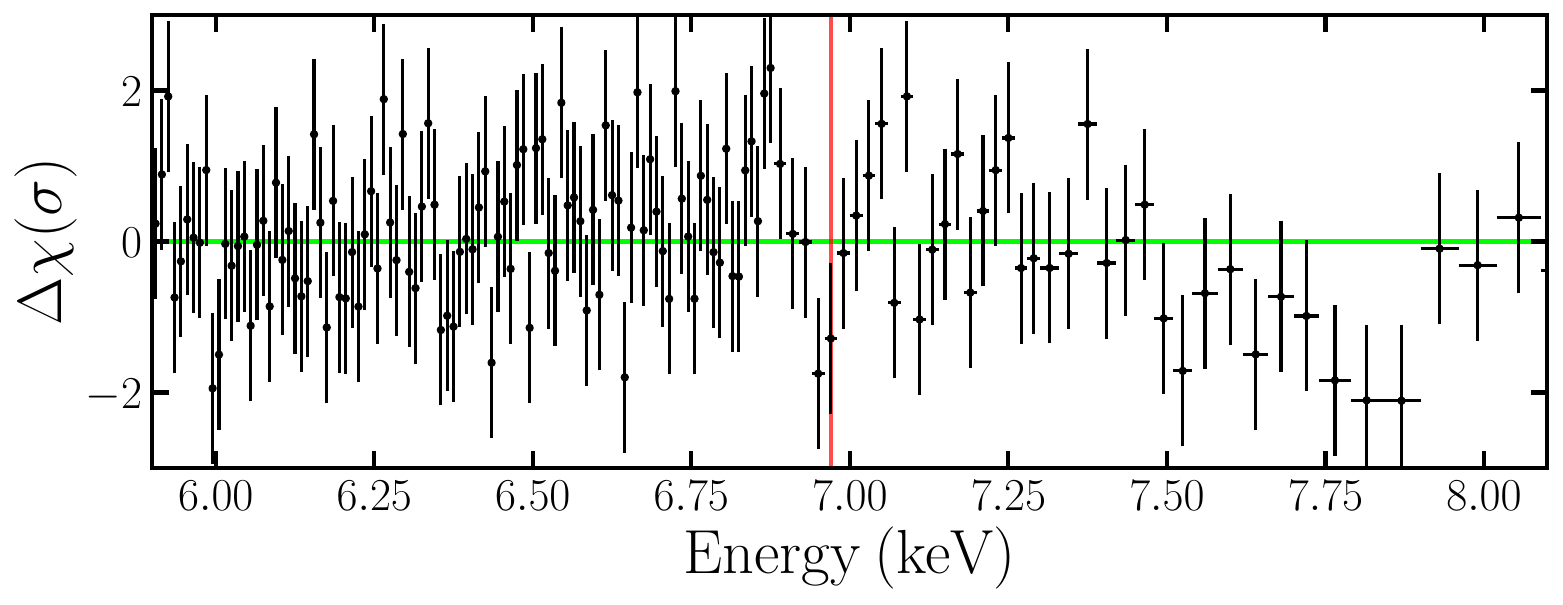}
    \caption{Residuals in terms of $\sigma$ in the 6-8 keV band produced when fitting the spectrum obtained by combining 52 NICER observations from the soft state of XTE J2012+381 using the best-fit reflection model obtained by fitting the second NuSTAR observation. The NICER observations used to obtain this combined spectrum are indicated by the two vertical dashed magenta lines in Figure \ref{fig:nicer_lc}. The vertical red line in this plot indicates the rest energy of the H-like Fe XXVI at 6.97 keV.}
    \label{fig:combined_residuals}
\end{figure}

While not apparent in the residuals produced when fitting the observations treated in this work, we tested whether a wind-like feature is present in this system by combining 52 NICER observations taken while the source was in a relatively stable soft state, throughout which we do not expect the continuum and reflection features to vary significantly. The observations used when combining the spectra are highlighted between the vertical dashed magenta lines in Figure \ref{fig:nicer_lc}. We used the \texttt{addspec.py} code written by Johannes Buchner \footnote{\url{https://github.com/JohannesBuchner/addspec.py}} to obtain an observation with an effective exposure of $\sim$190 ks, and the associated background and response files. The residuals produced when fitting the combined NICER spectrum with the best performing reflection model are shown in Figure \ref{fig:combined_residuals}. The red vertical line shows the position of the H-like Fe XXVI transition at 6.97~keV. While visually it appears that the residuals are suggestive of a narrow absorption-like feature, this is not statistically significant, similarly to the broader absorption-like feature just below 8 keV. Even if present, such a feature is unlikely to significantly impact the ability of the models to constrain the parameters, and the values obtained are likely to be impacted by degeneracies in the parameter space given the quality of the data. 

%paragraph about using kerrbb to estimate the mass of the BH
Lastly, we further explored the \texttt{kerrbb} variants of our models with the goal of placing better constraints on the mass of the BH in XTE J2012+381. We analyzed the best-fit models using the \texttt{kerrbb} component to describe the thermal emission from the accretion disk and the \texttt{relxill} component to describe the coronal emission and the reflected component, and used the models to perform joint fits of the NuSTAR and NICER observations during the two epochs. We linked the BH spin and inner disk inclination between the two components, fixed the normalization of the \texttt{kerrbb} component to 1, constrained the distance to the system to be between 3.3 kpc and 7.5 kpc as suggested by the Gaia measurement, and allowed all the other parameters of \texttt{kerrbb} to vary freely, namely the BH mass, the accretion rate, and the spectral hardening factor. Based on the observations from the first epoch, the mass is poorly constrained, giving $M=9.6^{+40.0}_{-1.5}\;\rm M_{\odot}$. However, the observations from the second epoch during which the disk component dominates over the coronal and reflected emission produces a BH mass constraint of $M=10.0^{+3.0}_{-0.4}\;\rm M_{\odot}$. It is important to note that for this measurement, there is a strong correlation between this parameter and the other parameters of the \texttt{kerrbb} component, and that the spectral hardening factor during the fit takes a very low value of $f\sim1$. Fits with larger values of $f$ fail to find similarly good solutions in terms of statistic for the observations during this epoch. 

\section{Discussion} \label{sec:disc}

%paragraph about the analysis and results
We analyzed two NuSTAR and 105 NICER observations of the late 2022 outburst of XTE J2012+381. By combining the information from two sets of simultaneous NICER and NuSTAR observations taken during two epochs three weeks apart, we measured the spin of the BH in the system to be $a=0.988^{+0.008}_{-0.030}$ and the inclination of the inner accretion disk to be $\theta=68^{+6}_{-11}$ degrees. This measurement was conducted using the pipeline established by \cite{2023ApJ...946...19D}, by testing an array of models describing the effects of relativistic reflection on the spectra, distinguishing the models using the DIC computed using the posterior distribution of an MCMC analysis, and combining the posterior distributions of the spin and inclination parameters using a Bayesian framework to maximize the information provided by all the existing observations. 

%paragraph about the importance of low-energy coverage, of obtaining multiple observations, and of observing as many times as possible to understand systematics
We ran our analysis pipeline on the two NuSTAR observations alone, and on joint fits to the NuSTAR spectra and simultaneous NICER spectra. When not including the low-energy coverage of NICER spectra, the measured Galactic column density is underestimated and the accretion disk temperature is slightly overestimated. In our analysis, we tested the effects of modeling the disk component both through the simplistic \texttt{diskbb} model or through the more physically accurate \texttt{kerrbb} model. Also, we tested the effects of including low-energy coverage through NICER spectra. Regardless of how the thermal emission from the accretion disk is modeled, as long as the continuum is well modeled, the reflection models are able to recover the shape of the relativistically broadened features and place agreeing constraints on the BH spin and viewing inclination of the inner accretion disk. 

This measurement highlights the importance of obtaining multiple observations when trying to understand the systematic uncertainties of BH spin measurements using relativistic reflection. The two main sources of systematic uncertainties for BH spin measurements come from peculiarities in the data and from aspects of the models that are not yet fully understood and characterized. An example of the former would be more complex phenomena than what we account for using our models, but which do not contribute significantly enough to be obviously required during the analysis (e.g. weak disk winds). An example of the latter is the effect of the accretion disk density in our models, and how that connects to the inferred Fe abundance, ionization, reflection fraction, and how that entirely contributes to our ability to constrain the underlying continuum to isolate the effects of reflection and measure the dynamic contributions on the broadening of spectral features, which directly constrain the spin.

In this work, we take the method used in \cite{2023ApJ...946...19D} to combine the posterior distributions of the spin and inclination parameters obtained from the MCMC analysis of the fits to independent NuSTAR observations and we expand it to better encapsulate and account for the systematic differences between independent measurements on different observations. As seen with the two observations of this source, obtaining stronger BH spin constraints is facilitated by having stronger reflection features during observations. Spectra taken while the sources are in harder states, where reflection is both stronger and easier to disentangle from the underlying continuum usually lead to better constraints on the BH spin than spectra taken during softer states. However, in order to understand the possible systematic differences that can lead to measurement uncertainties, multiple observations are required. Furthermore, obtaining multiple observations throughout the duration of BH XB outbursts reduces the likelihood of obtaining a single observation that does not allow placing reliable constraints on the BH spin (i.e., fitting only the NuSTAR observation 80802344004).

%paragraph about implications for spin distribution and connection to GW
The high spin of the BH in XTE J2012+381 is in good agreement with the distribution of spins measured in XB systems, and inconsistent with the distribution of spins of BHs in BBH mergers observed through GW. By expanding the observed sample of BH spins in XB, we begin to better explore possible observational biases that could explain the difference between the spins of BH in XB and the spins of BH in BBH. 

The distribution of BH spins can be used to construct a unified view of stellar-mass BH formation and evolution in binary systems. While high-mass X-ray binary (HMXB) systems are ideal candidates to link the population of XB to that of BBH as they contain a BH and a massive star that could also evolve to produce a secondary BH, \cite{2022ApJ...938L..19G} find that only up to 11\% of HMXB that experience an accretion episode while both stars are still on the main sequence (Case-A mass transfer) can evolve to eventually form a merging BBH system, and that at most 20\% of merging BBH systems originate from Case-A HMXB. Additionally, \cite{2023ApJ...946....4L} find that observational selection effects can further divide the link between HMXB and BBH through the fact that only around 0.6\% of detectable HMXB could produce a BBH system that would merge in a Hubble time. Therefore, independently understanding the different BH spin distributions is imperative, and the most pragmatic way to expand the spin distribution in XB is to continue to measure the spins of as many BHs as possible. 

%concluding paragraph about the future, talking about new outbursts and about XRISM and HEX-P
In the future, observations with XRISM (\citealt{2018SPIE10699E..22T}) and ATHENA (\citealt{2018SPIE10699E..1GB}) will provide high-resolution studies of the emission from XB systems, enabling more precise studies of the relativistically reflected radiation while better accounting for the effects of accretion disk winds, stellar companion winds, and the specifics of the physics of the accretion disk and the compact corona. Furthermore, missions such as HEX-P (\citealt{2018SPIE10699E..6MM}) or AXIS (\citealt{2018SPIE10699E..29M}) will be able to detect outbursts from more, fainter XB systems, significantly expanding the sample of measured BH spins in XB.

\begin{acknowledgments}
We thank the NuSTAR director, Fiona Harrison, and the mission scheduling team for making the observations. This research has made use of data and software provided by the High-Energy Astrophysics Science Archive Research Center (HEASARC), which is a service of the Astrophysics Science Division at NASA/GSFC, and of the NuSTAR Data Analysis Software (NuSTARDAS) jointly developed by the ASI Science Data Center (ASDC, Italy) and the California Institute of Technology (Caltech, USA). We thank the anonymous reviewer for their comments and suggestions, which have improved the quality of this paper.

\textit{Software:} 
\texttt{Astropy} (\citealt{astropy:2013, astropy:2018}), 
\texttt{emcee} (\citealt{2013PASP..125..306F}), 
\texttt{numpy} (\citealt{harris2020array}), 
\texttt{matplotlib} (\citealt{Hunter:2007}), 
\texttt{scipy} (\citealt{2020SciPy-NMeth}), 
\texttt{pandas} (\citealt{2022zndo...6702671R, mckinney-proc-scipy-2010}), 
\texttt{corner} (\citealt{corner}), 
\texttt{iPython} (\citealt{PER-GRA:2007}), 
\texttt{Xspec} (\citealt{1996ASPC..101...17A}), 
\texttt{relxill} (\citealt{2014MNRAS.444L.100D, 2014ApJ...782...76G}).
\end{acknowledgments}

\newpage

\appendix

\section{Corner plot}\label{sec:corner}

\begin{figure*}[ht!]
    \centering
    \includegraphics[width= 0.8\textwidth]{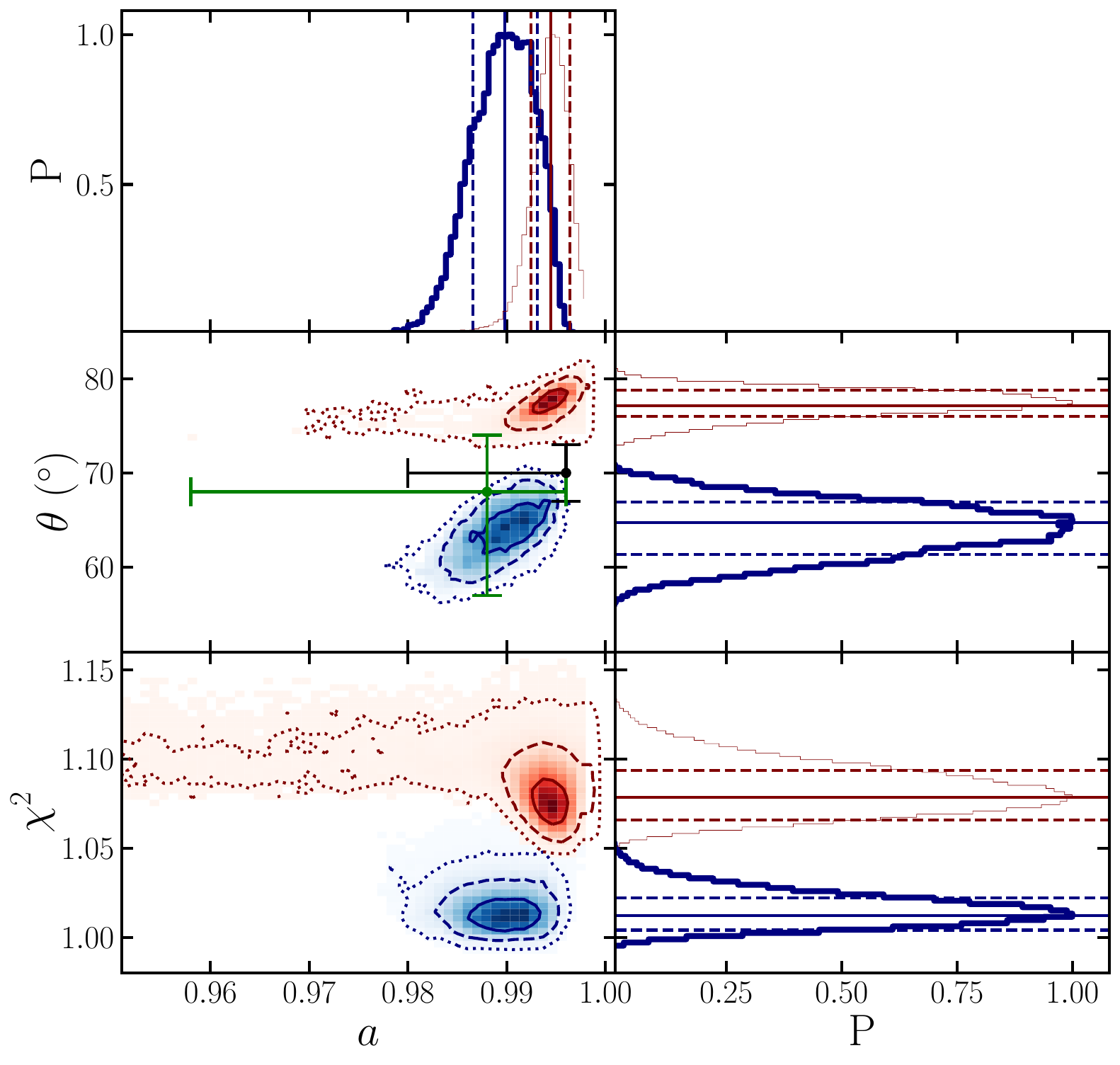}
    \caption{Two-dimensional histograms of the $a$-$\theta$ (center-left panel) and of the $a$-$\chi^2$ (bottom-left panel) parameter space based on the posterior samples in the MCMC analysis. The dashed, dash-dot, and dotted contours in these panels represent the $1\sigma$, $2\sigma$, and $3\sigma$ confidence intervals, respectively. The top-left, middle-right, and bottom-right panels show the 1D histograms of the posterior distributions in the MCMC analysis for spin, inclination, and reduced $\chi^2$. The width of the contours is proportional to the strength of reflection in the observation, which was used as weighting when combining the posterior distributions into a single measurement. The solid lines represent the modes of the distributions, and the dashed lines represent the $\pm1\sigma$ credible regions. Throughout the entire figure, the blue lines represent the results from the first epoch, and the red lines represent the results from the second epoch. In the middle-left panel, the black and green points show the values obtained by combining the posterior distributions through the two different methods highlighted in this paper.}
    \label{fig:spin_incl_chi}
\end{figure*}

The top-left, middle-right, and bottom-right panels in Figure \ref{fig:spin_incl_chi} show the 1D histograms of the posterior distributions of the spin, inclination, and reduced $\chi^2$ based on the MCMC analysis on the joint NuSTAR-NICER observations, with the blue curves representing the results from the first epoch, and the red curves representing the results from the second epoch. The width of the contours representing the histograms is proportional to the strength of reflection during the observations, which was used as weighting when comparing the independent measurements (see Figure \ref{fig:combined} and its explanation). The solid and dashed lines represent the modes and $1\sigma$ credible intervals of the individual posterior distributions. The histograms were normalized so that the peak of the distribution has a value of 1. The center-left and bottom-left panels show the 2D histograms of the $a$-$\theta$ and $a$-$\chi^2$ parameter space in the MCMC analysis, respectively. The dashed, dash-dot, and dotted contours in these panels represent the $1\sigma$, $2\sigma$, and $3\sigma$ confidence intervals, respectively. The black and green points represent the modes and uncertainty of the values obtained when combining the individual posterior distributions into a single distribution using the two methods highlighted in Figure \ref{fig:combined}. We report the green point as the result of this analysis.

\begin{figure*}[ht!]
    \centering
    \includegraphics[width= 0.8\textwidth]{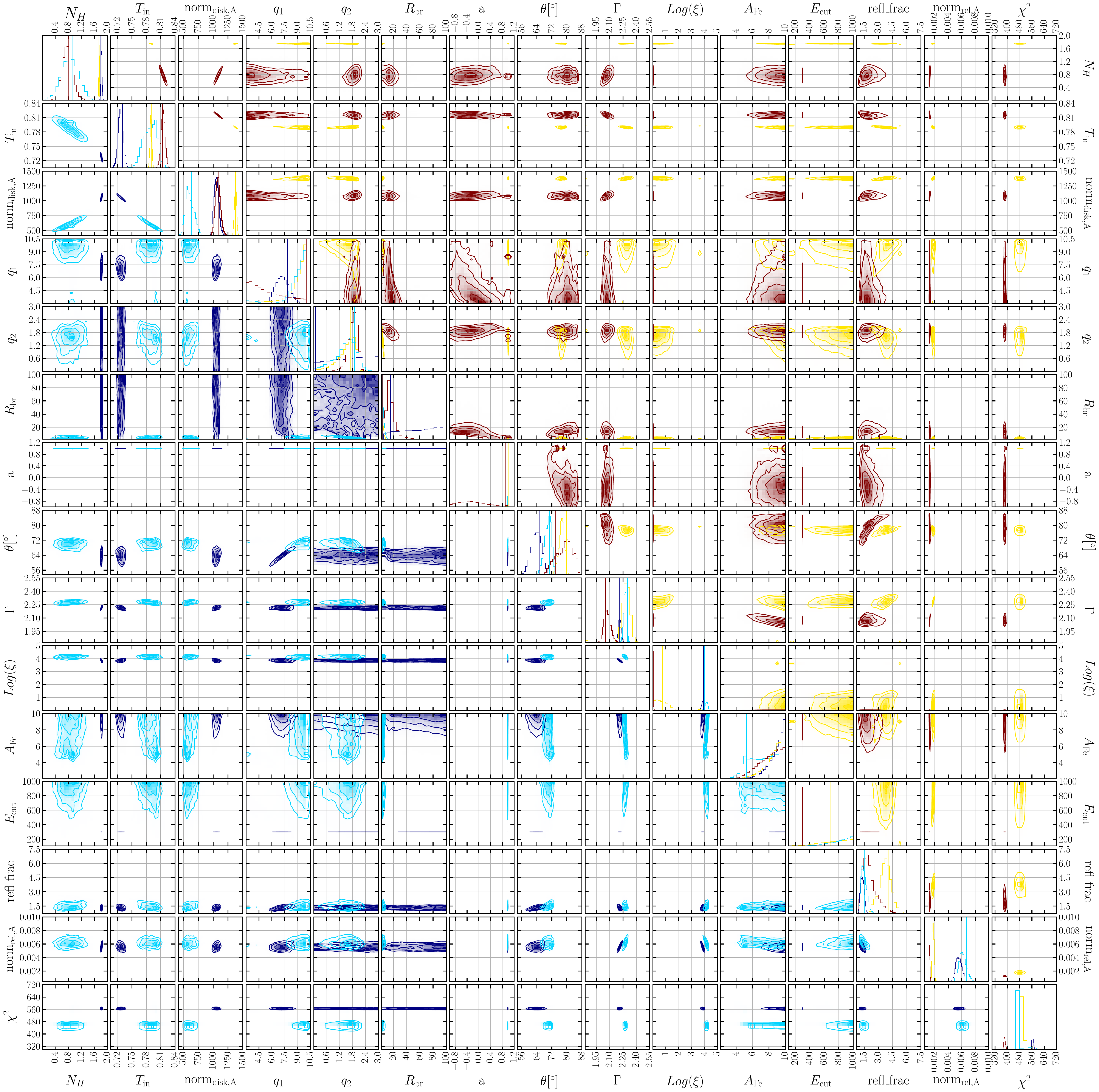}
    \caption{The complete corner plot of the MCMC analysis. The lower-left half of the plot shows the corner plot of the analysis of the observations from the first epoch, with the dark blue contours representing the joint NICER and NuSTAR fits, while the light blue ones representing the NuSTAR fits alone. The lop-right half of the plot shows the corner plot of the analysis of the observations from the second epoch, with the yellow contours showing the results from the joint NICER and NuSTAR fits, and the red ones showing the individual NuSTAR results. The sub-panels on the diagonal of the plot represent the 1D histograms of the marginalized posterior distributions for each parameter in the MCMC analysis. The vertical lines in the diagonal sub-plots represent the values around which Gaussian proposal distributions were generated and used to initialize the walkers in the MCMC run. }
    \label{fig:corner}
\end{figure*}

Figure \ref{fig:corner} shows the complete corner plot generated from the posterior samples resulting from the MCMC analysis. The diagonal entries show the marginalized 1-dimensional probability distributions for the individual parameters in the analysis, and the rest of the panels show the 2-dimensional regions of the parameter space for combinations of parameters in the models. The bottom-left half of the plot shows the corner plot of the observations from the first epoch, with the dark blue contours representing the results from the joint NICER and NuSTAR analysis, and the light blue representing the results from the independent NuSTAR analysis. The top-right corner shows the corner plot of the analysis from the second epoch, with the yellow contours showing the results of the joint NICER and NuSTAR analysis, and the red contours showing the individual NuSTAR analysis. The contours in the plot represent the $1\sigma$, $2\sigma$, and $3\sigma$ confidence intervals in the 2D posterior distribution for each parameter combination. The vertical lines in the 1D posterior distributions (the subplots on the diagonal) represent the values around which Gaussian proposal distributions were generated and used to initialize the walkers in the MCMC run. The priors for the parameters are uniform in the parameter range allowed by the model components. For simplicity, we only plot the normalizations of the \texttt{diskbb} and \texttt{relxill} components for the NuSTAR FPMA spectra.

The most obvious trends noticeable are the correlations between the Galactic column density $N_{\rm H}$ and the parameters of the \texttt{diskbb} model, namely the disk temperature $kT_{\rm in}$ and the normalization of the component $norm_{\rm disk,A}$. However, as discussed in above, as long as the continuum is fully characterized, the reflection features are identified to be the same, producing very similar spin measurements. The data are often unable to properly constrain the outer emissivity index $q_2$ and the breaking radius $R_{\rm br}$, as the emissivity is often steep in the inner disk regions, strongly suppressing the contribution at larger distances. As discussed in the main text, the inner emissivity index contributes strongly to the ability to constrain the BH spin. As we can see here, the outer emissivity and the breaking radius do not impact the spin directly, but rather through the ability to constrain $q_1$, which in turn influences our ability to measure the inner disk radius which constrains the spin.

\newpage
\bibliography{paper}{}

\begin{thebibliography}{}
\expandafter\ifx\csname natexlab\endcsname\relax\def\natexlab#1{#1}\fi
\providecommand{\url}[1]{\href{#1}{#1}}
\providecommand{\dodoi}[1]{doi:~\href{http://doi.org/#1}{\nolinkurl{#1}}}
\providecommand{\doeprint}[1]{\href{http://ascl.net/#1}{\nolinkurl{http://ascl.net/#1}}}
\providecommand{\doarXiv}[1]{\href{https://arxiv.org/abs/#1}{\nolinkurl{https://arxiv.org/abs/#1}}}

\bibitem[{{Arnaud}(1996)}]{1996ASPC..101...17A}
{Arnaud}, K.~A. 1996, in Astronomical Society of the Pacific Conference Series,
  Vol. 101, Astronomical Data Analysis Software and Systems V, ed. G.~H.
  {Jacoby} \& J.~{Barnes}, 17

\bibitem[{{Astropy Collaboration} {et~al.}(2013){Astropy Collaboration},
  {Robitaille}, {Tollerud}, {Greenfield}, {Droettboom}, {Bray}, {Aldcroft},
  {Davis}, {Ginsburg}, {Price-Whelan}, {Kerzendorf}, {Conley}, {Crighton},
  {Barbary}, {Muna}, {Ferguson}, {Grollier}, {Parikh}, {Nair}, {Unther},
  {Deil}, {Woillez}, {Conseil}, {Kramer}, {Turner}, {Singer}, {Fox}, {Weaver},
  {Zabalza}, {Edwards}, {Azalee Bostroem}, {Burke}, {Casey}, {Crawford},
  {Dencheva}, {Ely}, {Jenness}, {Labrie}, {Lim}, {Pierfederici}, {Pontzen},
  {Ptak}, {Refsdal}, {Servillat}, \& {Streicher}}]{astropy:2013}
{Astropy Collaboration}, {Robitaille}, T.~P., {Tollerud}, E.~J., {et~al.} 2013,
  \aap, 558, A33, \dodoi{10.1051/0004-6361/201322068}

\bibitem[{{Astropy Collaboration} {et~al.}(2018){Astropy Collaboration},
  {Price-Whelan}, {Sip{\H{o}}cz}, {G{\"u}nther}, {Lim}, {Crawford}, {Conseil},
  {Shupe}, {Craig}, {Dencheva}, {Ginsburg}, {VanderPlas}, {Bradley},
  {P{\'e}rez-Su{\'a}rez}, {de Val-Borro}, {Aldcroft}, {Cruz}, {Robitaille},
  {Tollerud}, {Ardelean}, {Babej}, {Bach}, {Bachetti}, {Bakanov}, {Bamford},
  {Barentsen}, {Barmby}, {Baumbach}, {Berry}, {Biscani}, {Boquien}, {Bostroem},
  {Bouma}, {Brammer}, {Bray}, {Breytenbach}, {Buddelmeijer}, {Burke},
  {Calderone}, {Cano Rodr{\'\i}guez}, {Cara}, {Cardoso}, {Cheedella}, {Copin},
  {Corrales}, {Crichton}, {D'Avella}, {Deil}, {Depagne}, {Dietrich}, {Donath},
  {Droettboom}, {Earl}, {Erben}, {Fabbro}, {Ferreira}, {Finethy}, {Fox},
  {Garrison}, {Gibbons}, {Goldstein}, {Gommers}, {Greco}, {Greenfield},
  {Groener}, {Grollier}, {Hagen}, {Hirst}, {Homeier}, {Horton}, {Hosseinzadeh},
  {Hu}, {Hunkeler}, {Ivezi{\'c}}, {Jain}, {Jenness}, {Kanarek}, {Kendrew},
  {Kern}, {Kerzendorf}, {Khvalko}, {King}, {Kirkby}, {Kulkarni}, {Kumar},
  {Lee}, {Lenz}, {Littlefair}, {Ma}, {Macleod}, {Mastropietro}, {McCully},
  {Montagnac}, {Morris}, {Mueller}, {Mumford}, {Muna}, {Murphy}, {Nelson},
  {Nguyen}, {Ninan}, {N{\"o}the}, {Ogaz}, {Oh}, {Parejko}, {Parley}, {Pascual},
  {Patil}, {Patil}, {Plunkett}, {Prochaska}, {Rastogi}, {Reddy Janga},
  {Sabater}, {Sakurikar}, {Seifert}, {Sherbert}, {Sherwood-Taylor}, {Shih},
  {Sick}, {Silbiger}, {Singanamalla}, {Singer}, {Sladen}, {Sooley},
  {Sornarajah}, {Streicher}, {Teuben}, {Thomas}, {Tremblay}, {Turner},
  {Terr{\'o}n}, {van Kerkwijk}, {de la Vega}, {Watkins}, {Weaver}, {Whitmore},
  {Woillez}, {Zabalza}, \& {Astropy Contributors}}]{astropy:2018}
{Astropy Collaboration}, {Price-Whelan}, A.~M., {Sip{\H{o}}cz}, B.~M., {et~al.}
  2018, \aj, 156, 123, \dodoi{10.3847/1538-3881/aabc4f}

\bibitem[{{Barret} {et~al.}(2018){Barret}, {Lam Trong}, {den Herder}, \&
  et~al.}]{2018SPIE10699E..1GB}
{Barret}, D., {Lam Trong}, T., {den Herder}, J.-W., \& et~al. 2018, in Society
  of Photo-Optical Instrumentation Engineers (SPIE) Conference Series, Vol.
  10699, Space Telescopes and Instrumentation 2018: Ultraviolet to Gamma Ray,
  ed. J.-W.~A. {den Herder}, S.~{Nikzad}, \& K.~{Nakazawa}, 106991G,
  \dodoi{10.1117/12.2312409}

\bibitem[{{Brenneman} \& {Reynolds}(2006)}]{2006ApJ...652.1028B}
{Brenneman}, L.~W., \& {Reynolds}, C.~S. 2006, ApJ, 652, 1028,
  \dodoi{10.1086/508146}

\bibitem[{{Campana} {et~al.}(2002){Campana}, {Stella}, {Belloni}, {Israel},
  {Santangelo}, {Frontera}, {Orlandini}, \& {Dal Fiume}}]{2002A&A...384..163C}
{Campana}, S., {Stella}, L., {Belloni}, T., {et~al.} 2002, \aap, 384, 163,
  \dodoi{10.1051/0004-6361:20020012}

\bibitem[{{Connors} {et~al.}(2021){Connors}, {Garc{\'\i}a}, {Tomsick}, {Hare},
  {Dauser}, {Grinberg}, {Steiner}, {Mastroserio}, {Sridhar}, {Fabian}, {Jiang},
  {Parker}, {Harrison}, \& {Kallman}}]{2021ApJ...909..146C}
{Connors}, R. M.~T., {Garc{\'\i}a}, J.~A., {Tomsick}, J., {et~al.} 2021, \apj,
  909, 146, \dodoi{10.3847/1538-4357/abdd2c}

\bibitem[{{Dauser} {et~al.}(2014){Dauser}, {Garcia}, {Parker}, {Fabian}, \&
  {Wilms}}]{2014MNRAS.444L.100D}
{Dauser}, T., {Garcia}, J., {Parker}, M.~L., {Fabian}, A.~C., \& {Wilms}, J.
  2014, MNRAS, 444, L100, \dodoi{10.1093/mnrasl/slu125}

\bibitem[{{Dauser} {et~al.}(2022){Dauser}, {Garc{\'\i}a}, {Joyce},
  {Licklederer}, {Connors}, {Ingram}, {Reynolds}, \&
  {Wilms}}]{2022MNRAS.514.3965D}
{Dauser}, T., {Garc{\'\i}a}, J.~A., {Joyce}, A., {et~al.} 2022, \mnras, 514,
  3965, \dodoi{10.1093/mnras/stac1593}

\bibitem[{{Draghis} {et~al.}(2020){Draghis}, {Miller}, {Cackett}, {Kammoun},
  {Reynolds}, {Tomsick}, \& {Zoghbi}}]{2020ApJ...900...78D}
{Draghis}, P.~A., {Miller}, J.~M., {Cackett}, E.~M., {et~al.} 2020, \apj, 900,
  78, \dodoi{10.3847/1538-4357/aba2ec}

\bibitem[{{Draghis} {et~al.}(2021){Draghis}, {Miller}, {Zoghbi}, {Kammoun},
  {Reynolds}, \& {Tomsick}}]{2021ApJ...920...88D}
{Draghis}, P.~A., {Miller}, J.~M., {Zoghbi}, A., {et~al.} 2021, \apj, 920, 88,
  \dodoi{10.3847/1538-4357/ac1270}

\bibitem[{{Draghis} {et~al.}(2023{\natexlab{a}}){Draghis}, {Miller}, {Zoghbi},
  {Reynolds}, {Costantini}, {Gallo}, \& {Tomsick}}]{2023ApJ...946...19D}
---. 2023{\natexlab{a}}, \apj, 946, 19, \dodoi{10.3847/1538-4357/acafe7}

\bibitem[{{Draghis} {et~al.}(2023{\natexlab{b}}){Draghis}, {Balakrishnan},
  {Miller}, {Cackett}, {Fabian}, {Miller-Jones}, {Ng}, {Raymond}, {Reynolds},
  \& {Zoghbi}}]{2023ApJ...947...39D}
{Draghis}, P.~A., {Balakrishnan}, M., {Miller}, J.~M., {et~al.}
  2023{\natexlab{b}}, \apj, 947, 39, \dodoi{10.3847/1538-4357/acc1c8}

\bibitem[{{Fabian} {et~al.}(2000){Fabian}, {Iwasawa}, {Reynolds}, \&
  {Young}}]{2000PASP..112.1145F}
{Fabian}, A.~C., {Iwasawa}, K., {Reynolds}, C.~S., \& {Young}, A.~J. 2000,
  \pasp, 112, 1145, \dodoi{10.1086/316610}

\bibitem[{{Fabian} {et~al.}(2014){Fabian}, {Parker}, {Wilkins}, {Miller},
  {Kara}, {Reynolds}, \& {Dauser}}]{2014MNRAS.439.2307F}
{Fabian}, A.~C., {Parker}, M.~L., {Wilkins}, D.~R., {et~al.} 2014, \mnras, 439,
  2307, \dodoi{10.1093/mnras/stu045}

\bibitem[{{Feng} {et~al.}(2023){Feng}, {Steiner}, {Ramirez}, \&
  {Gou}}]{2023MNRAS.520.5803F}
{Feng}, Y., {Steiner}, J.~F., {Ramirez}, S.~U., \& {Gou}, L. 2023, \mnras, 520,
  5803, \dodoi{10.1093/mnras/stad442}

\bibitem[{{Fishbach} \& {Kalogera}(2022)}]{2022ApJ...929L..26F}
{Fishbach}, M., \& {Kalogera}, V. 2022, \apjl, 929, L26,
  \dodoi{10.3847/2041-8213/ac64a5}

\bibitem[{Foreman-Mackey(2016)}]{corner}
Foreman-Mackey, D. 2016, The Journal of Open Source Software, 1, 24,
  \dodoi{10.21105/joss.00024}

\bibitem[{{Foreman-Mackey} {et~al.}(2013){Foreman-Mackey}, {Hogg}, {Lang}, \&
  {Goodman}}]{2013PASP..125..306F}
{Foreman-Mackey}, D., {Hogg}, D.~W., {Lang}, D., \& {Goodman}, J. 2013, \pasp,
  125, 306, \dodoi{10.1086/670067}

\bibitem[{{Gaia Collaboration} {et~al.}(2016){Gaia Collaboration}, {Prusti},
  {de Bruijne}, {Brown}, {Vallenari}, {Babusiaux}, {Bailer-Jones}, {Bastian},
  {Biermann}, {Evans}, {Eyer}, {Jansen}, {Jordi}, {Klioner}, {Lammers},
  {Lindegren}, {Luri}, {Mignard}, {Milligan}, {Panem}, {Poinsignon},
  {Pourbaix}, {Randich}, {Sarri}, {Sartoretti}, {Siddiqui}, {Soubiran},
  {Valette}, {van Leeuwen}, {Walton}, {Aerts}, {Arenou}, {Cropper}, {Drimmel},
  {H{\o}g}, {Katz}, {Lattanzi}, {O'Mullane}, {Grebel}, {Holland}, {Huc},
  {Passot}, {Bramante}, {Cacciari}, {Casta{\~n}eda}, {Chaoul}, {Cheek}, {De
  Angeli}, {Fabricius}, {Guerra}, {Hern{\'a}ndez}, {Jean-Antoine-Piccolo},
  {Masana}, {Messineo}, {Mowlavi}, {Nienartowicz}, {Ord{\'o}{\~n}ez-Blanco},
  {Panuzzo}, {Portell}, {Richards}, {Riello}, {Seabroke}, {Tanga},
  {Th{\'e}venin}, {Torra}, {Els}, {Gracia-Abril}, {Comoretto},
  {Garcia-Reinaldos}, {Lock}, {Mercier}, {Altmann}, {Andrae}, {Astraatmadja},
  {Bellas-Velidis}, {Benson}, {Berthier}, {Blomme}, {Busso}, {Carry},
  {Cellino}, {Clementini}, {Cowell}, {Creevey}, {Cuypers}, {Davidson}, {De
  Ridder}, {de Torres}, {Delchambre}, {Dell'Oro}, {Ducourant}, {Fr{\'e}mat},
  {Garc{\'\i}a-Torres}, {Gosset}, {Halbwachs}, {Hambly}, {Harrison}, {Hauser},
  {Hestroffer}, {Hodgkin}, {Huckle}, {Hutton}, {Jasniewicz}, {Jordan},
  {Kontizas}, {Korn}, {Lanzafame}, {Manteiga}, {Moitinho}, {Muinonen},
  {Osinde}, {Pancino}, {Pauwels}, {Petit}, {Recio-Blanco}, {Robin}, {Sarro},
  {Siopis}, {Smith}, {Smith}, {Sozzetti}, {Thuillot}, {van Reeven}, {Viala},
  {Abbas}, {Abreu Aramburu}, {Accart}, {Aguado}, {Allan}, {Allasia},
  {Altavilla}, {{\'A}lvarez}, {Alves}, {Anderson}, {Andrei}, {Anglada Varela},
  {Antiche}, {Antoja}, {Ant{\'o}n}, {Arcay}, {Atzei}, {Ayache}, {Bach},
  {Baker}, {Balaguer-N{\'u}{\~n}ez}, {Barache}, {Barata}, {Barbier}, {Barblan},
  {Baroni}, {Barrado y Navascu{\'e}s}, {Barros}, {Barstow}, {Becciani},
  {Bellazzini}, {Bellei}, {Bello Garc{\'\i}a}, {Belokurov}, {Bendjoya},
  {Berihuete}, {Bianchi}, {Bienaym{\'e}}, {Billebaud}, {Blagorodnova},
  {Blanco-Cuaresma}, {Boch}, {Bombrun}, {Borrachero}, {Bouquillon}, {Bourda},
  {Bouy}, {Bragaglia}, {Breddels}, {Brouillet}, {Br{\"u}semeister},
  {Bucciarelli}, {Budnik}, {Burgess}, {Burgon}, {Burlacu}, {Busonero}, {Buzzi},
  {Caffau}, {Cambras}, {Campbell}, {Cancelliere}, {Cantat-Gaudin}, {Carlucci},
  {Carrasco}, {Castellani}, {Charlot}, {Charnas}, {Charvet}, {Chassat},
  {Chiavassa}, {Clotet}, {Cocozza}, {Collins}, {Collins}, {Costigan}, {Crifo},
  {Cross}, {Crosta}, {Crowley}, {Dafonte}, {Damerdji}, {Dapergolas}, {David},
  {David}, {De Cat}, {de Felice}, {de Laverny}, {De Luise}, {De March}, {de
  Martino}, {de Souza}, {Debosscher}, {del Pozo}, {Delbo}, {Delgado},
  {Delgado}, {di Marco}, {Di Matteo}, {Diakite}, {Distefano}, {Dolding}, {Dos
  Anjos}, {Drazinos}, {Dur{\'a}n}, {Dzigan}, {Ecale}, {Edvardsson}, {Enke},
  {Erdmann}, {Escolar}, {Espina}, {Evans}, {Eynard Bontemps}, {Fabre},
  {Fabrizio}, {Faigler}, {Falc{\~a}o}, {Farr{\`a}s Casas}, {Faye}, {Federici},
  {Fedorets}, {Fern{\'a}ndez-Hern{\'a}ndez}, {Fernique}, {Fienga}, {Figueras},
  {Filippi}, {Findeisen}, {Fonti}, {Fouesneau}, {Fraile}, {Fraser}, {Fuchs},
  {Furnell}, {Gai}, {Galleti}, {Galluccio}, {Garabato}, {Garc{\'\i}a-Sedano},
  {Gar{\'e}}, {Garofalo}, {Garralda}, {Gavras}, {Gerssen}, {Geyer}, {Gilmore},
  {Girona}, {Giuffrida}, {Gomes}, {Gonz{\'a}lez-Marcos},
  {Gonz{\'a}lez-N{\'u}{\~n}ez}, {Gonz{\'a}lez-Vidal}, {Granvik}, {Guerrier},
  {Guillout}, {Guiraud}, {G{\'u}rpide}, {Guti{\'e}rrez-S{\'a}nchez}, {Guy},
  {Haigron}, {Hatzidimitriou}, {Haywood}, {Heiter}, {Helmi}, {Hobbs},
  {Hofmann}, {Holl}, {Holland}, {Hunt}, {Hypki}, {Icardi}, {Irwin}, {Jevardat
  de Fombelle}, {Jofr{\'e}}, {Jonker}, {Jorissen}, {Julbe}, {Karampelas},
  {Kochoska}, {Kohley}, {Kolenberg}, {Kontizas}, {Koposov}, {Kordopatis},
  {Koubsky}, {Kowalczyk}, {Krone-Martins}, {Kudryashova}, {Kull}, {Bachchan},
  {Lacoste-Seris}, {Lanza}, {Lavigne}, {Le Poncin-Lafitte}, {Lebreton},
  {Lebzelter}, {Leccia}, {Leclerc}, {Lecoeur-Taibi}, {Lemaitre}, {Lenhardt},
  {Leroux}, {Liao}, {Licata}, {Lindstr{\o}m}, {Lister}, {Livanou}, {Lobel},
  {L{\"o}ffler}, {L{\'o}pez}, {Lopez-Lozano}, {Lorenz}, {Loureiro},
  {MacDonald}, {Magalh{\~a}es Fernandes}, {Managau}, {Mann}, {Mantelet},
  {Marchal}, {Marchant}, {Marconi}, {Marie}, {Marinoni}, {Marrese},
  {Marschalk{\'o}}, {Marshall}, {Mart{\'\i}n-Fleitas}, {Martino}, {Mary},
  {Matijevi{\v{c}}}, {Mazeh}, {McMillan}, {Messina}, {Mestre}, {Michalik},
  {Millar}, {Miranda}, {Molina}, {Molinaro}, {Molinaro}, {Moln{\'a}r},
  {Moniez}, {Montegriffo}, {Monteiro}, {Mor}, {Mora}, {Morbidelli}, {Morel},
  {Morgenthaler}, {Morley}, {Morris}, {Mulone}, {Muraveva}, {Musella},
  {Narbonne}, {Nelemans}, {Nicastro}, {Noval}, {Ord{\'e}novic},
  {Ordieres-Mer{\'e}}, {Osborne}, {Pagani}, {Pagano}, {Pailler}, {Palacin},
  {Palaversa}, {Parsons}, {Paulsen}, {Pecoraro}, {Pedrosa}, {Pentik{\"a}inen},
  {Pereira}, {Pichon}, {Piersimoni}, {Pineau}, {Plachy}, {Plum}, {Poujoulet},
  {Pr{\v{s}}a}, {Pulone}, {Ragaini}, {Rago}, {Rambaux}, {Ramos-Lerate},
  {Ranalli}, {Rauw}, {Read}, {Regibo}, {Renk}, {Reyl{\'e}}, {Ribeiro},
  {Rimoldini}, {Ripepi}, {Riva}, {Rixon}, {Roelens}, {Romero-G{\'o}mez},
  {Rowell}, {Royer}, {Rudolph}, {Ruiz-Dern}, {Sadowski}, {Sagrist{\`a}
  Sell{\'e}s}, {Sahlmann}, {Salgado}, {Salguero}, {Sarasso}, {Savietto},
  {Schnorhk}, {Schultheis}, {Sciacca}, {Segol}, {Segovia}, {Segransan},
  {Serpell}, {Shih}, {Smareglia}, {Smart}, {Smith}, {Solano}, {Solitro},
  {Sordo}, {Soria Nieto}, {Souchay}, {Spagna}, {Spoto}, {Stampa}, {Steele},
  {Steidelm{\"u}ller}, {Stephenson}, {Stoev}, {Suess}, {S{\"u}veges}, {Surdej},
  {Szabados}, {Szegedi-Elek}, {Tapiador}, {Taris}, {Tauran}, {Taylor},
  {Teixeira}, {Terrett}, {Tingley}, {Trager}, {Turon}, {Ulla}, {Utrilla},
  {Valentini}, {van Elteren}, {Van Hemelryck}, {van Leeuwen}, {Varadi},
  {Vecchiato}, {Veljanoski}, {Via}, {Vicente}, {Vogt}, {Voss}, {Votruba},
  {Voutsinas}, {Walmsley}, {Weiler}, {Weingrill}, {Werner}, {Wevers},
  {Whitehead}, {Wyrzykowski}, {Yoldas}, {{\v{Z}}erjal}, {Zucker}, {Zurbach},
  {Zwitter}, {Alecu}, {Allen}, {Allende Prieto}, {Amorim},
  {Anglada-Escud{\'e}}, {Arsenijevic}, {Azaz}, {Balm}, {Beck}, {Bernstein},
  {Bigot}, {Bijaoui}, {Blasco}, {Bonfigli}, {Bono}, {Boudreault}, {Bressan},
  {Brown}, {Brunet}, {Bunclark}, {Buonanno}, {Butkevich}, {Carret}, {Carrion},
  {Chemin}, {Ch{\'e}reau}, {Corcione}, {Darmigny}, {de Boer}, {de Teodoro}, {de
  Zeeuw}, {Delle Luche}, {Domingues}, {Dubath}, {Fodor}, {Fr{\'e}zouls},
  {Fries}, {Fustes}, {Fyfe}, {Gallardo}, {Gallegos}, {Gardiol}, {Gebran},
  {Gomboc}, {G{\'o}mez}, {Grux}, {Gueguen}, {Heyrovsky}, {Hoar}, {Iannicola},
  {Isasi Parache}, {Janotto}, {Joliet}, {Jonckheere}, {Keil}, {Kim},
  {Klagyivik}, {Klar}, {Knude}, {Kochukhov}, {Kolka}, {Kos}, {Kutka}, {Lainey},
  {LeBouquin}, {Liu}, {Loreggia}, {Makarov}, {Marseille}, {Martayan},
  {Martinez-Rubi}, {Massart}, {Meynadier}, {Mignot}, {Munari}, {Nguyen},
  {Nordlander}, {Ocvirk}, {O'Flaherty}, {Olias Sanz}, {Ortiz}, {Osorio},
  {Oszkiewicz}, {Ouzounis}, {Palmer}, {Park}, {Pasquato}, {Peltzer}, {Peralta},
  {P{\'e}turaud}, {Pieniluoma}, {Pigozzi}, {Poels}, {Prat}, {Prod'homme},
  {Raison}, {Rebordao}, {Risquez}, {Rocca-Volmerange}, {Rosen}, {Ruiz-Fuertes},
  {Russo}, {Sembay}, {Serraller Vizcaino}, {Short}, {Siebert}, {Silva},
  {Sinachopoulos}, {Slezak}, {Soffel}, {Sosnowska}, {Strai{\v{z}}ys}, {ter
  Linden}, {Terrell}, {Theil}, {Tiede}, {Troisi}, {Tsalmantza}, {Tur},
  {Vaccari}, {Vachier}, {Valles}, {Van Hamme}, {Veltz}, {Virtanen}, {Wallut},
  {Wichmann}, {Wilkinson}, {Ziaeepour}, \& {Zschocke}}]{2016A&A...595A...1G}
{Gaia Collaboration}, {Prusti}, T., {de Bruijne}, J.~H.~J., {et~al.} 2016,
  \aap, 595, A1, \dodoi{10.1051/0004-6361/201629272}

\bibitem[{{Gallegos-Garcia} {et~al.}(2022){Gallegos-Garcia}, {Fishbach},
  {Kalogera}, {L Berry}, \& {Doctor}}]{2022ApJ...938L..19G}
{Gallegos-Garcia}, M., {Fishbach}, M., {Kalogera}, V., {L Berry}, C.~P., \&
  {Doctor}, Z. 2022, \apjl, 938, L19, \dodoi{10.3847/2041-8213/ac96ef}

\bibitem[{{Garc{\'\i}a} {et~al.}(2014){Garc{\'\i}a}, {Dauser}, {Lohfink},
  {Kallman}, {Steiner}, {McClintock}, {Brenneman}, {Wilms}, {Eikmann},
  {Reynolds}, \& {Tombesi}}]{2014ApJ...782...76G}
{Garc{\'\i}a}, J., {Dauser}, T., {Lohfink}, A., {et~al.} 2014, ApJ, 782, 76,
  \dodoi{10.1088/0004-637X/782/2/76}

\bibitem[{{Garc{\'\i}a} {et~al.}(2015){Garc{\'\i}a}, {Steiner}, {McClintock},
  {Remillard}, {Grinberg}, \& {Dauser}}]{2015ApJ...813...84G}
{Garc{\'\i}a}, J.~A., {Steiner}, J.~F., {McClintock}, J.~E., {et~al.} 2015,
  \apj, 813, 84, \dodoi{10.1088/0004-637X/813/2/84}

\bibitem[{{Gendreau} {et~al.}(2016){Gendreau}, {Arzoumanian}, {Adkins},
  {Albert}, {Anders}, {Aylward}, {Baker}, {Balsamo}, {Bamford}, {Benegalrao},
  {Berry}, {Bhalwani}, {Black}, {Blaurock}, {Bronke}, {Brown}, {Budinoff},
  {Cantwell}, {Cazeau}, {Chen}, {Clement}, {Colangelo}, {Coleman},
  {Coopersmith}, {Dehaven}, {Doty}, {Egan}, {Enoto}, {Fan}, {Ferro}, {Foster},
  {Galassi}, {Gallo}, {Green}, {Grosh}, {Ha}, {Hasouneh}, {Heefner}, {Hestnes},
  {Hoge}, {Jacobs}, {J{\o}rgensen}, {Kaiser}, {Kellogg}, {Kenyon}, {Koenecke},
  {Kozon}, {LaMarr}, {Lambertson}, {Larson}, {Lentine}, {Lewis}, {Lilly},
  {Liu}, {Malonis}, {Manthripragada}, {Markwardt}, {Matonak}, {Mcginnis},
  {Miller}, {Mitchell}, {Mitchell}, {Mohammed}, {Monroe}, {Montt de Garcia},
  {Mul{\'e}}, {Nagao}, {Ngo}, {Norris}, {Norwood}, {Novotka}, {Okajima},
  {Olsen}, {Onyeachu}, {Orosco}, {Peterson}, {Pevear}, {Pham}, {Pollard},
  {Pope}, {Powers}, {Powers}, {Price}, {Prigozhin}, {Ramirez}, {Reid},
  {Remillard}, {Rogstad}, {Rosecrans}, {Rowe}, {Sager}, {Sanders}, {Savadkin},
  {Saylor}, {Schaeffer}, {Schweiss}, {Semper}, {Serlemitsos}, {Shackelford},
  {Soong}, {Struebel}, {Vezie}, {Villasenor}, {Winternitz}, {Wofford},
  {Wright}, {Yang}, \& {Yu}}]{2016SPIE.9905E..1HG}
{Gendreau}, K.~C., {Arzoumanian}, Z., {Adkins}, P.~W., {et~al.} 2016, in
  Society of Photo-Optical Instrumentation Engineers (SPIE) Conference Series,
  Vol. 9905, Space Telescopes and Instrumentation 2016: Ultraviolet to Gamma
  Ray, ed. J.-W.~A. {den Herder}, T.~{Takahashi}, \& M.~{Bautz}, 99051H,
  \dodoi{10.1117/12.2231304}

\bibitem[{{Gou} {et~al.}(2009){Gou}, {McClintock}, {Liu}, {Narayan}, {Steiner},
  {Remillard}, {Orosz}, {Davis}, {Ebisawa}, \&
  {Schlegel}}]{2009ApJ...701.1076G}
{Gou}, L., {McClintock}, J.~E., {Liu}, J., {et~al.} 2009, \apj, 701, 1076,
  \dodoi{10.1088/0004-637X/701/2/1076}

\bibitem[{Harris {et~al.}(2020)Harris, Millman, van~der Walt, Gommers,
  Virtanen, Cournapeau, Wieser, Taylor, Berg, Smith, Kern, Picus, Hoyer, van
  Kerkwijk, Brett, Haldane, del R{\'{i}}o, Wiebe, Peterson,
  G{\'{e}}rard-Marchant, Sheppard, Reddy, Weckesser, Abbasi, Gohlke, \&
  Oliphant}]{harris2020array}
Harris, C.~R., Millman, K.~J., van~der Walt, S.~J., {et~al.} 2020, Nature, 585,
  357, \dodoi{10.1038/s41586-020-2649-2}

\bibitem[{{Harrison} {et~al.}(2013){Harrison}, {Craig}, {Christensen}, \&
  et~al.}]{2013ApJ...770..103H}
{Harrison}, F.~A., {Craig}, W.~W., {Christensen}, F.~E., \& et~al. 2013, ApJ,
  770, 103, \dodoi{10.1088/0004-637X/770/2/103}

\bibitem[{Hunter(2007)}]{Hunter:2007}
Hunter, J.~D. 2007, Computing in Science \& Engineering, 9, 90,
  \dodoi{10.1109/MCSE.2007.55}

\bibitem[{{Hynes} {et~al.}(1999){Hynes}, {Roche}, {Charles}, \&
  {Coe}}]{1999MNRAS.305L..49H}
{Hynes}, R.~I., {Roche}, P., {Charles}, P.~A., \& {Coe}, M.~J. 1999, \mnras,
  305, L49, \dodoi{10.1046/j.1365-8711.1999.02653.x}

\bibitem[{{Jiang} {et~al.}(2020){Jiang}, {Gallo}, {Fabian}, {Parker}, \&
  {Reynolds}}]{2020MNRAS.498.3888J}
{Jiang}, J., {Gallo}, L.~C., {Fabian}, A.~C., {Parker}, M.~L., \& {Reynolds},
  C.~S. 2020, \mnras, 498, 3888, \dodoi{10.1093/mnras/staa2625}

\bibitem[{{Kaastra} \& {Bleeker}(2016)}]{2016A&A...587A.151K}
{Kaastra}, J.~S., \& {Bleeker}, J.~A.~M. 2016, \aap, 587, A151,
  \dodoi{10.1051/0004-6361/201527395}

\bibitem[{{Kawamuro} {et~al.}(2022){Kawamuro}, {Negoro}, {Nakajima},
  {Kobayashi}, {Tanaka}, {Soejima}, {Mihara}, {Yamada}, {Tamagawa}, {Matsuoka},
  {Sakamoto}, {Serino}, {Sugita}, {Hiramatsu}, {Nishikawa}, {Yoshida},
  {Tsuboi}, {Kohara}, {Urabe}, {Nawa}, {Nemoto}, {Iwakiri}, {Shidatsu},
  {Iwasaki}, {Kawai}, {Niwano}, {Hosokawa}, {Imai}, {Ito}, {Takamatsu},
  {Nakahira}, {Ueno}, {Tomida}, {Ishikawa}, {Kurihara}, {Ueda}, {Ogawa},
  {Setoguchi}, {Yoshitake}, {Inaba}, {Yamauchi}, {Sato}, {Hatsuda}, {Fukuoka},
  {Hagiwara}, {Umeki}, {Yamaoka}, {Kawakubo}, \&
  {Sugizaki}}]{2022ATel15826....1K}
{Kawamuro}, T., {Negoro}, H., {Nakajima}, M., {et~al.} 2022, The Astronomer's
  Telegram, 15826, 1

\bibitem[{{Kennea}(2022)}]{2022ATel15827....1K}
{Kennea}, J.~A. 2022, The Astronomer's Telegram, 15827, 1

\bibitem[{{King} {et~al.}(2012){King}, {Miller}, {Raymond}, {Fabian},
  {Reynolds}, {Kallman}, {Maitra}, {Cackett}, \& {Rupen}}]{2012ApJ...746L..20K}
{King}, A.~L., {Miller}, J.~M., {Raymond}, J., {et~al.} 2012, \apjl, 746, L20,
  \dodoi{10.1088/2041-8205/746/2/L20}

\bibitem[{{King} {et~al.}(2014){King}, {Walton}, {Miller}, {Barret}, {Boggs},
  {Christensen}, {Craig}, {Fabian}, {F{\"u}rst}, {Hailey}, {Harrison},
  {Krivonos}, {Mori}, {Natalucci}, {Stern}, {Tomsick}, \&
  {Zhang}}]{2014ApJ...784L...2K}
{King}, A.~L., {Walton}, D.~J., {Miller}, J.~M., {et~al.} 2014, \apjl, 784, L2,
  \dodoi{10.1088/2041-8205/784/1/L2}

\bibitem[{{Li} {et~al.}(2005){Li}, {Zimmerman}, {Narayan}, \&
  {McClintock}}]{2005ApJS..157..335L}
{Li}, L.-X., {Zimmerman}, E.~R., {Narayan}, R., \& {McClintock}, J.~E. 2005,
  ApJs, 157, 335, \dodoi{10.1086/428089}

\bibitem[{{Liotine} {et~al.}(2023){Liotine}, {Zevin}, {Berry}, {Doctor}, \&
  {Kalogera}}]{2023ApJ...946....4L}
{Liotine}, C., {Zevin}, M., {Berry}, C. P.~L., {Doctor}, Z., \& {Kalogera}, V.
  2023, \apj, 946, 4, \dodoi{10.3847/1538-4357/acb8b2}

\bibitem[{{Madsen} {et~al.}(2018){Madsen}, {Harrison}, {Broadway},
  {Christensen}, {Descalle}, {Ferreira}, {Grefenstette}, {Gurgew},
  {Hornschemeier}, {Miyasaka}, {Okajima}, {Pike}, {Pivovaroff}, {Saha},
  {Stern}, {Vogel}, {Windt}, \& {Zhang}}]{2018SPIE10699E..6MM}
{Madsen}, K.~K., {Harrison}, F., {Broadway}, D., {et~al.} 2018, in Society of
  Photo-Optical Instrumentation Engineers (SPIE) Conference Series, Vol. 10699,
  Space Telescopes and Instrumentation 2018: Ultraviolet to Gamma Ray, ed.
  J.-W.~A. {den Herder}, S.~{Nikzad}, \& K.~{Nakazawa}, 106996M,
  \dodoi{10.1117/12.2314117}

\bibitem[{{Matt} {et~al.}(1993){Matt}, {Fabian}, \&
  {Ross}}]{1993MNRAS.262..179M}
{Matt}, G., {Fabian}, A.~C., \& {Ross}, R.~R. 1993, \mnras, 262, 179,
  \dodoi{10.1093/mnras/262.1.179}

\bibitem[{{Miller}(2007)}]{2007ARA&A..45..441M}
{Miller}, J.~M. 2007, \araa, 45, 441,
  \dodoi{10.1146/annurev.astro.45.051806.110555}

\bibitem[{{Miller} {et~al.}(2006){Miller}, {Raymond}, {Homan}, {Fabian},
  {Steeghs}, {Wijnands}, {Rupen}, {Charles}, {van der Klis}, \&
  {Lewin}}]{2006ApJ...646..394M}
{Miller}, J.~M., {Raymond}, J., {Homan}, J., {et~al.} 2006, \apj, 646, 394,
  \dodoi{10.1086/504673}

\bibitem[{{Miller} {et~al.}(2010){Miller}, {D'A{\`\i}}, {Bautz},
  {Bhattacharyya}, {Burrows}, {Cackett}, {Fabian}, {Freyberg}, {Haberl},
  {Kennea}, {Nowak}, {Reis}, {Strohmayer}, \&
  {Tsujimoto}}]{2010ApJ...724.1441M}
{Miller}, J.~M., {D'A{\`\i}}, A., {Bautz}, M.~W., {et~al.} 2010, \apj, 724,
  1441, \dodoi{10.1088/0004-637X/724/2/1441}

\bibitem[{{Mushotzky}(2018)}]{2018SPIE10699E..29M}
{Mushotzky}, R. 2018, in Society of Photo-Optical Instrumentation Engineers
  (SPIE) Conference Series, Vol. 10699, Space Telescopes and Instrumentation
  2018: Ultraviolet to Gamma Ray, ed. J.-W.~A. {den Herder}, S.~{Nikzad}, \&
  K.~{Nakazawa}, 1069929, \dodoi{10.1117/12.2310003}

\bibitem[{P\'erez \& Granger(2007)}]{PER-GRA:2007}
P\'erez, F., \& Granger, B.~E. 2007, Computing in Science and Engineering, 9,
  21, \dodoi{10.1109/MCSE.2007.53}

\bibitem[{{Ponti} {et~al.}(2012){Ponti}, {Fender}, {Begelman}, {Dunn},
  {Neilsen}, \& {Coriat}}]{2012MNRAS.422L..11P}
{Ponti}, G., {Fender}, R.~P., {Begelman}, M.~C., {et~al.} 2012, \mnras, 422,
  L11, \dodoi{10.1111/j.1745-3933.2012.01224.x}

\bibitem[{{Reback} {et~al.}(2022){Reback}, {Jbrockmendel}, {McKinney}, {Van Den
  Bossche}, {Roeschke}, {Augspurger}, {Hawkins}, {Cloud}, {Gfyoung}, {Sinhrks},
  {Hoefler}, {Klein}, {Petersen}, {Tratner}, {She}, {Ayd}, {Naveh},
  {Darbyshire}, {Shadrach}, {Garcia}, {Schendel}, {Hayden}, {Saxton},
  {Gorelli}, {Li}, {W{\"o}rtwein}, {Zeitlin}, {Jancauskas}, {McMaster}, \&
  {Li}}]{2022zndo...6702671R}
{Reback}, J., {Jbrockmendel}, {McKinney}, W., {et~al.} 2022,
  {pandas-dev/pandas: Pandas 1.4.3}, v1.4.3, Zenodo,  Zenodo,
  \dodoi{10.5281/zenodo.6702671}

\bibitem[{{Remillard} {et~al.}(1998){Remillard}, {Levine}, {Wood}, {Wagner},
  {Starrfield}, {Shrader}, {Bowell}, {Skiff}, \& {Koehn}}]{1998IAUC.6920....1R}
{Remillard}, R., {Levine}, A., {Wood}, A., {et~al.} 1998, \iaucirc, 6920, 1

\bibitem[{{Reynolds}(2021)}]{2021ARA&A..59..117R}
{Reynolds}, C.~S. 2021, \araa, 59, 117,
  \dodoi{10.1146/annurev-astro-112420-035022}

\bibitem[{{Reynolds} {et~al.}(2012){Reynolds}, {Brenneman}, {Lohfink},
  {Trippe}, {Miller}, {Fabian}, \& {Nowak}}]{2012ApJ...755...88R}
{Reynolds}, C.~S., {Brenneman}, L.~W., {Lohfink}, A.~M., {et~al.} 2012, \apj,
  755, 88, \dodoi{10.1088/0004-637X/755/2/88}

\bibitem[{{Reynolds} \& {Fabian}(2008)}]{2008ApJ...675.1048R}
{Reynolds}, C.~S., \& {Fabian}, A.~C. 2008, \apj, 675, 1048,
  \dodoi{10.1086/527344}

\bibitem[{{Riaz} {et~al.}(2023){Riaz}, {Abdikamalov}, \&
  {Bambi}}]{2023arXiv230312581R}
{Riaz}, S., {Abdikamalov}, A.~B., \& {Bambi}, C. 2023, arXiv e-prints,
  arXiv:2303.12581, \dodoi{10.48550/arXiv.2303.12581}

\bibitem[{{Rodriguez} {et~al.}(2023){Rodriguez}, {Chenevez}, {Tomsick},
  {Wilms}, {Fiocchi}, {Bazzano}, {Pottschmidt}, {Steiner}, {Bouchet},
  {Cangemi}, {Coleiro}, {Egron}, {Grinberg}, {Petrucci}, \&
  {Clavel}}]{2023ATel15847....1R}
{Rodriguez}, J., {Chenevez}, J., {Tomsick}, J., {et~al.} 2023, The Astronomer's
  Telegram, 15847, 1

\bibitem[{{Salvesen} {et~al.}(2013){Salvesen}, {Miller}, {Reis}, \&
  {Begelman}}]{2013MNRAS.431.3510S}
{Salvesen}, G., {Miller}, J.~M., {Reis}, R.~C., \& {Begelman}, M.~C. 2013,
  MNRAS, 431, 3510, \dodoi{10.1093/mnras/stt436}

\bibitem[{{Schnittman} {et~al.}(2016){Schnittman}, {Krolik}, \&
  {Noble}}]{2016ApJ...819...48S}
{Schnittman}, J.~D., {Krolik}, J.~H., \& {Noble}, S.~C. 2016, ApJ, 819, 48,
  \dodoi{10.3847/0004-637X/819/1/48}

\bibitem[{Spiegelhalter {et~al.}(2002)Spiegelhalter, Best, Carlin, \& Van
  Der~Linde}]{DIC_text}
Spiegelhalter, D.~J., Best, N.~G., Carlin, B.~P., \& Van Der~Linde, A. 2002,
  Journal of the Royal Statistical Society: Series B (Statistical Methodology),
  64, 583, \dodoi{https://doi.org/10.1111/1467-9868.00353}

\bibitem[{{Tashiro} {et~al.}(2018){Tashiro}, {Maejima}, {Toda}, \&
  et~al.}]{2018SPIE10699E..22T}
{Tashiro}, M., {Maejima}, H., {Toda}, K., \& et~al. 2018, in Space Telescopes
  and Instrumentation 2018: Ultraviolet to Gamma Ray, ed. J.-W.~A. den Herder,
  S.~Nikzad, \& K.~Nakazawa, Vol. 10699, International Society for Optics and
  Photonics (SPIE), 520 -- 531, \dodoi{10.1117/12.2309455}

\bibitem[{{Tomsick} {et~al.}(2018){Tomsick}, {Parker}, {Garc{\'\i}a},
  {Yamaoka}, {Barret}, {Chiu}, {Clavel}, {Fabian}, {F{\"u}rst}, {Gandhi},
  {Grinberg}, {Miller}, {Pottschmidt}, \& {Walton}}]{2018ApJ...855....3T}
{Tomsick}, J.~A., {Parker}, M.~L., {Garc{\'\i}a}, J.~A., {et~al.} 2018, \apj,
  855, 3, \dodoi{10.3847/1538-4357/aaaab1}

\bibitem[{{Vasiliev} {et~al.}(2000){Vasiliev}, {Trudolyubov}, \&
  {Revnivtsev}}]{2000A&A...362L..53V}
{Vasiliev}, L., {Trudolyubov}, S., \& {Revnivtsev}, M. 2000, \aap, 362, L53,
  \dodoi{10.48550/arXiv.astro-ph/0008176}

\bibitem[{{Verner} {et~al.}(1996){Verner}, {Ferland}, {Korista}, \&
  {Yakovlev}}]{1996ApJ...465..487V}
{Verner}, D.~A., {Ferland}, G.~J., {Korista}, K.~T., \& {Yakovlev}, D.~G. 1996,
  ApJ, 465, 487, \dodoi{10.1086/177435}

\bibitem[{Virtanen {et~al.}(2020)Virtanen, Gommers, Oliphant, Haberland, Reddy,
  Cournapeau, Burovski, Peterson, Weckesser, Bright, {van der Walt}, Brett,
  Wilson, Millman, Mayorov, Nelson, Jones, Kern, Larson, Carey, Polat, Feng,
  Moore, {VanderPlas}, Laxalde, Perktold, Cimrman, Henriksen, Quintero, Harris,
  Archibald, Ribeiro, Pedregosa, {van Mulbregt}, \& {SciPy 1.0
  Contributors}}]{2020SciPy-NMeth}
Virtanen, P., Gommers, R., Oliphant, T.~E., {et~al.} 2020, Nature Methods, 17,
  261, \dodoi{10.1038/s41592-019-0686-2}

\bibitem[{{W}es {M}c{K}inney(2010)}]{mckinney-proc-scipy-2010}
{W}es {M}c{K}inney. 2010, in {P}roceedings of the 9th {P}ython in {S}cience
  {C}onference, ed. {S}t\'efan van~der {W}alt \& {J}arrod {M}illman, 56 -- 61,
  \dodoi{10.25080/Majora-92bf1922-00a}

\bibitem[{{White} {et~al.}(1998){White}, {Ueda}, {Dotani}, \&
  {Nagase}}]{1998IAUC.6927....2W}
{White}, N.~E., {Ueda}, Y., {Dotani}, T., \& {Nagase}, F. 1998, \iaucirc, 6927,
  2

\bibitem[{{Wilms} {et~al.}(2000){Wilms}, {Allen}, \&
  {McCray}}]{2000ApJ...542..914W}
{Wilms}, J., {Allen}, A., \& {McCray}, R. 2000, ApJ, 542, 914,
  \dodoi{10.1086/317016}

\bibitem[{{Wilms} {et~al.}(2006){Wilms}, {Juett}, {Schulz}, \&
  {Nowak}}]{2006HEAD....9.1360W}
{Wilms}, J., {Juett}, A., {Schulz}, N., \& {Nowak}, M. 2006, in AAS/High Energy
  Astrophysics Division \#9, AAS/High Energy Astrophysics Division, 13.60

\end{thebibliography}
\bibliographystyle{aasjournal}

\end{document}